\documentclass{ptephy_v1}

\preprintnumber{XXXX-XXXX} 

\usepackage{graphicx}
\usepackage{dcolumn}
\usepackage{bm}
\usepackage[nooneline]{subfigure}
\usepackage{amsmath}
\usepackage{mathtools}
\usepackage{bm}
\usepackage{mathrsfs}
\usepackage{comment}
\usepackage{natbib}

\begin{document}

\title{Simulation study on the effects of diffractive collisions on the prediction of the observables in ultra-high-energy cosmic ray experiments}


\author{Ken Ohashi }
\affil{Institute for Space-Earth Environmental Research, Nagoya University, Nagoya, Japan \email{ohashi.ken@isee.nagoya-u.ac.jp}}

\author[1]{Hiroaki Menjo}
\author[1,2]{Yoshitaka Itow}
\affil{Kobayashi-Maskawa Institute for the Origin of Particles and the Universe, Nagoya University, Nagoya, Japan}
\author[3]{Takashi Sako}
\affil{Institute for Cosmic Ray Research, the University of Tokyo, Kashiwa, Japan}

\author[4]{Katsuaki Kasahara}
\affil{College of Systems Engineering and Science, Shibaura Institute  of Technology, Minuma-ku, Saitama, Japan}



\begin{abstract}%
The mass composition of ultra-high-energy cosmic rays is important for understanding their origin. Owing to our limited knowledge of the hadronic interaction, the interpretations of the mass composition from observations include several open problems, such as the inconsistent interpretations of $\langle X_{\mathrm{max}}\rangle $ and $\langle X_{\mathrm{max}}^{\mu}\rangle $. Futhermore, the large difference between the predictions exists by the hadronic interaction models. Diffractive collision is one of the proposed sources of the uncertainty. In this paper, we discuss the effect of the detailed characteristics of diffractive collisions to the observables of ultra-high-energy cosmic-ray experiments, focusing on three detailed characteristics. These are the cross-sectional fractions of different collision types, diffractive-mass spectrum, and diffractive-mass-dependent particle productions from the diffractive dissociation system. We demonstrated that the current level of the uncertainty in the cross-sectional fraction can affect 8.9~$\mathrm{g/cm^2}$ of $\langle X_{\mathrm{max}}\rangle $ and 9.4~~$\mathrm{g/cm^2}$ of $\langle X_{\mathrm{max}}^{\mu}\rangle $, whereas the other details of the diffractive collisions exhibit relatively  minor effects.  
\end{abstract}

\subjectindex{xxxx, xxx}

\maketitle


\section{\label{sec:Intro}Introduction}

The origin of ultra-high-energy cosmic rays (UHECRs) above $10^{18}$~eV is one of the most important questions in astrophysics. The mass composition of these cosmic rays provides a key information on the understanding of their origin. UHECRs are observed by the measurements of the extended air showers. The mass-sensitive observables are extracted from the air shower data, such as the depth of the maximum of the shower development $X_{\mathrm{max}}$, the depth of the maximum of the muon productions in an air shower $X_{\mathrm{max}}^{\mu}$, and the number of muons at the ground $N_{\mu}$. 
The mass composition is estimated by comparing these observables, measured by the Pierre Auger Observatory~\cite{Aab2014,Aab2014a,Collica2016,Mallamaci_2017,Aab2015} and the Telescope Array Collaboration~\cite{TelescopeArrayCollaboration2018}, with predictions based on the Monte Carlo (MC) simulations. However, owing to the limited knowledge on the hadronic interactions at such a high energy, the interpretations of the mass composition have several problems. 
Actually, the interpretation from $X_{\mathrm{max}}^{\mu}$ predicts a heavier composition in comparison with the prediction from $X_{\mathrm{max}}$~\cite{Mallamaci_2017}, and the predictions on $N_{\mu}$ provide us much smaller value than the experimental data~\cite{Aab2015}. The excess of $N_{\mu}$ in the observations with respect to the MC predictions is widely recognized as the "muon excess problem". Recent experimental results related to this problem are summarized in Ref.~\cite{WorkingGroupReport}. 
In addition, there are significant differences between the predictions on the air shower observables calculated by different hadronic interaction models.
For example, the difference in the predictions on $\langle X_{\mathrm{max}}\rangle $, which is the average of $X_{\mathrm{max}}$, of the models for $10^{19}$~eV proton primary cosmic rays is 30~$\mathrm{g/cm^2}$, whereas the size of the systematic uncertainties in the experiments is 15~$\mathrm{g/cm^2}$ \cite{Aab2014}. As the $\langle X_{\mathrm{max}}\rangle $ predictions from simulations become larger, 
the interpretations of mass composition from the experimental data become heavier. 
Therefore, precise understanding and improvements of the treatments on the hadronic interactions in simulations are required.

In MC simulations, hadronic interaction models are used to simulate the interactions between a hadron and an air nucleus. Since quantum chromodynamics is difficult to study by the first-principle calculations for the low-momentum transfer processes, such calculations in the hadronic interaction models are described, being based on the phenomenology. Diffractive collisions are characterized by the low momentum transfer. 
Therefore, there are large uncertainties in the predictions of these collisions. Moreover, energies of the produced particles in diffractive collisions is larger than the other collisions, thus these collisions may be a large impact on the air shower development~\cite{ulrich2011}.

A few previous studies discussed the effect of diffractive collisions on the UHECR observables. 
Under an extreme assumption, turning on and off the diffractive collisions in the air-shower simulations, the difference in $\langle X_{\mathrm{max}}\rangle $ was estimated to be as large as 15~(14)~$\mathrm{g/cm^{2}}$ for $10^{20}$~eV proton (iron) primaries~\cite{ArbeletcheAirShower}.
In another study, The effect of the different modeling of the diffractive collisions was discussed by changing 
the differential cross section of diffractive-mass in SIBYLL~2.3c over whole air showers of $10^{19}$~eV proton primaries with a $67^{\circ}$-incident zenith angle~\cite{SIBYLL-diff}.
The effect turned out as to be $+5.0$~$\mathrm{g/cm^2}$ for $\langle X_{\mathrm{max}}\rangle $ and -5~\% for the average number of muons at the depth, $X=2240~\mathrm{g/cm^2}$.
Even though these analyses indicated that the effects of the diffractive collisions are not negligible, 
only the total effect and the effect of the differential cross section of diffractive-mass in one model are discussed in the previous works. The effects of different types of diffractive collisions, such as single, double, and central diffractions, and the effect of the diffractive mass distributions and particle productions from diffractive dissociation, 
remain unclear. 

In this study, to understand the impact of the detailed characteristics of diffractive collisions on the UHECR mass composition analyses, the effect of these characteristics on predicting the UHECR observables using the air shower simulation package, CONEX~v6.40~\cite{conex}, was studied. In particular, the composition-sensitive observables, $X_{\mathrm{max}}$ and $X_{\mathrm{max}}^{\mu}$, were focused.
For this purpose, several characteristics of the diffractive collision processes were parameterized. By changing these parameters within the variations in the different models, the possible effects on the observables were discussed.
The modifications were applied only for the first interaction, in case A, and over all the interactions in air showers, in case B. This case B modification is used for a study of the effects of cross-section fractions in Sec.~\ref{sec:XmaxFractionResampling}. The case A modification is used for all studies including the effects of cross-section fractions and the modeling of diffractive collisions in Sec.~\ref{sec:XmaxFractionFirstint} and \ref{sec:XmaxDiffMass}. 
The details of the diffractive collisions and the simulation set-up are discussed in Sec.~\ref{sec:Diffractive} and \ref{sec:AirShowerSimulation}, respectively. Basic profiles of longitudinal air shower developments are also discussed in Sec.~\ref{sec:AirShowerSimulation}. Subsequently, we discuss the effects on the longitudinal properties of air showers,  $\langle X_{\mathrm{max}}\rangle $ and $\langle X_{\mathrm{max}}^{\mu}\rangle $, in Sec.~\ref{sec:EffectOnXmax}. Then, the effects on the fluctuation of $X_{\mathrm{max}}$ is discussed in Sec.~\ref{sec:SigmaXmax}. 
The conclusion is provided in Sec.~\ref{sec:Conclusion}.

\section{\label{sec:Diffractive}Diffractive collisions}

\begin{figure}
    \centering
    \includegraphics[clip, width=0.9\columnwidth]{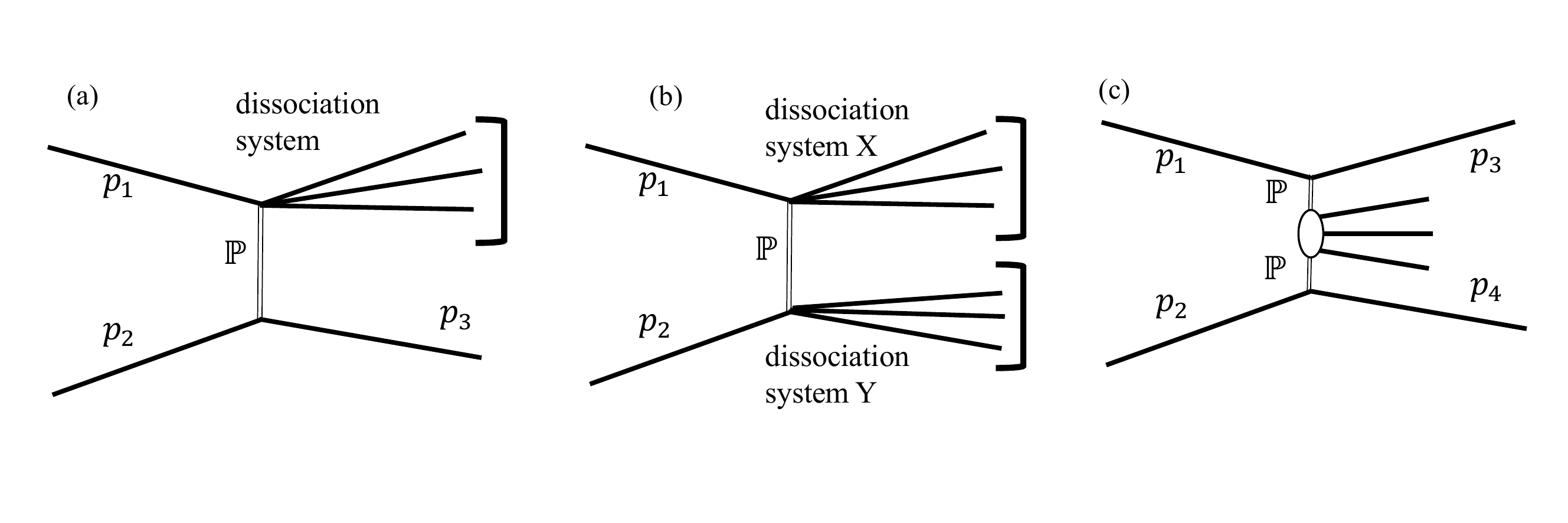}
    \caption{Feynman diagrams of (a) single diffractive (SD) collisions, (b) double diffractive (DD) collisions, and (c) central diffractive (CD) collisions. I$\!$P represents a pomeron.}
    \label{fig:Diagram}
\end{figure}

\begin{figure}
    \centering
        \includegraphics[clip, width=0.85\columnwidth]{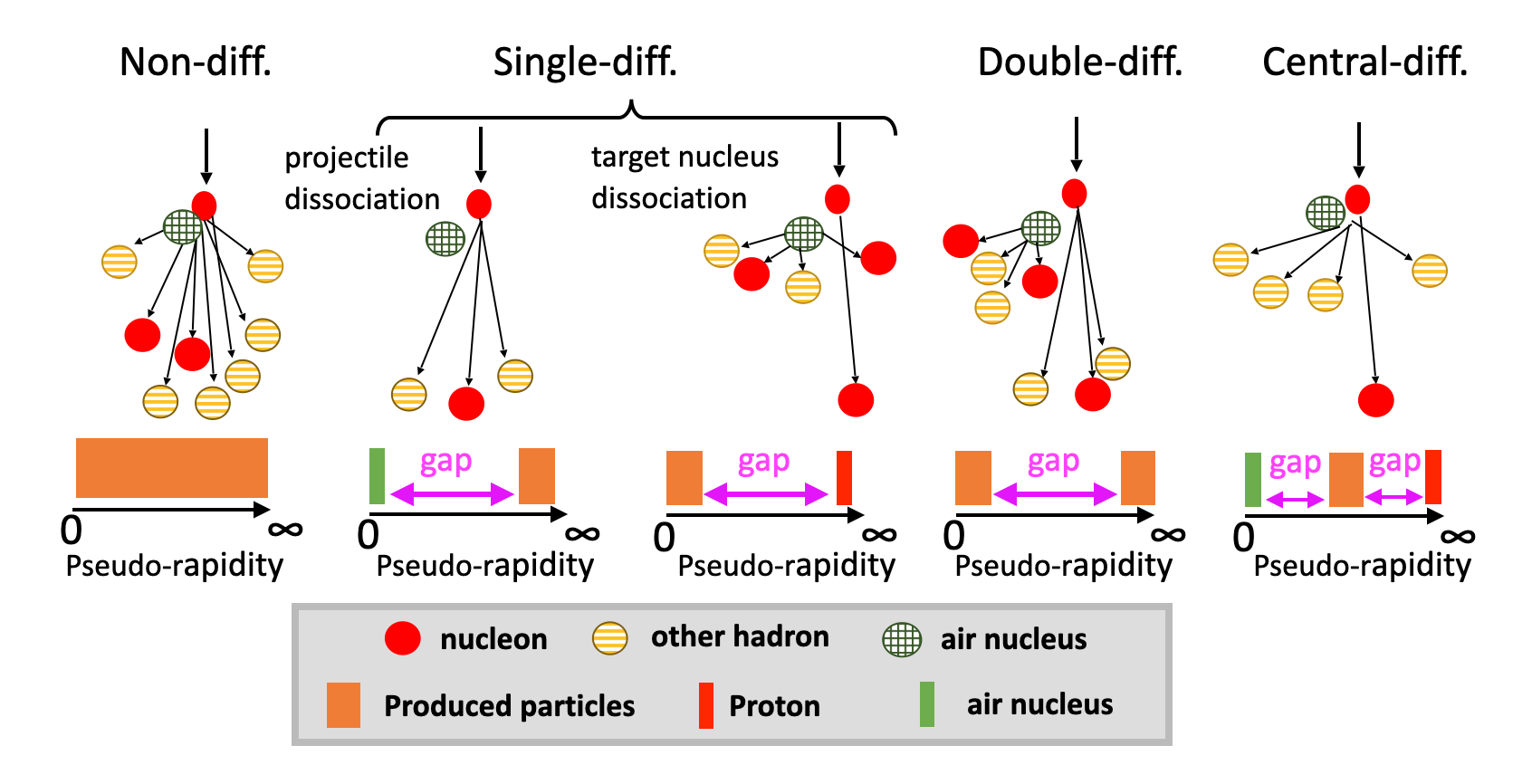}
    \caption{Schematic of the collision between a cosmic ray proton and an air nucleus for each collision type. Illustrations of rapidity distributions of produced particles for each collision type are also shown in the bottom side. }
    \label{fig:SchematicView}
\end{figure}

Diffractive collisions are explained using the exchange of a virtual particle without any quantum numbers called Pomeron. The mechanism of the diffractive collisions, including particle productions from the diffractive dissociation system, is not well understood yet. Experimentally, these collisions are characterized by a rapidity gap of the produced particles from the dissociation of the incident particles~\cite{LHCForwardPhysics}.
Because the energy of the produced particles in the diffractive collisions is expected to be higher than in the other types of collisions, the diffractive collision is expected to be more relevant to the air shower observables~\cite{ulrich2011}. 
There are three types of diffractive collisions: single diffractive (SD) collisions, double diffractive (DD) collisions, and central diffractive (CD) collisions. One of the initial particles dissociates, whereas the other is intact in the SD collisions. Both the initial particles dissociate in the DD collisions. The CD collisions are diffractive collisions with particle productions from collisions between two or more pomerons without dissociation of the initial particles, and these collisions typically have two rapidity gaps. 
Figure~\ref{fig:Diagram} displays the Feynman diagrams of these collisions. In this study, the inelastic collisions excluding the diffractive collisions are called as non-diffractive (ND) collisions. The SD collisions with a projectile cosmic-ray dissociation and those with a target air nucleus dissociation induce different effects on the air shower development, as displayed in Fig.~\ref{fig:SchematicView}. In this paper, the former collision type is called as the projectile single diffractive (pSD) collision, whereas the latter is called as the target single diffractive (tSD) collision. 

In single and double diffractive collisions, an incident particle dissociate to particles and the modeling of this dissociation process has large uncertainty. In hadronic interaction models, this dissociation process is modeled using the diffractive mass $M_{X}$; an incident particle is excited by the pomeron exchange and particles are produced from the dissociation of the excited state, for example the decay of a resonance in low diffractive mass region or the string fragmentation in high diffractive mass region. 
The modeling of the diffractive collision vary in hadronic interaction models, and they are summarized in Ref.\cite{LHCForwardPhysics}.
$M_{X}$ can be calculated using the momentum of the produced particles in a dissociation system as, 
\begin{equation}
    M_{X}^{2}=\left| \sum_{i}^{n} \bm{p}_{i} \right|^2,
    \label{eq:DiffMass}
\end{equation}
where $\bm{p}_{i}$ and $n$ are the four-momentum of the $i$-th particle and the number of particles in the dissociation system, respectively. The modeling of the diffractive collisions can be divided into two parts: the differential cross-section of diffractive mass, $\frac{d\sigma}{ dM_{X}}$, and treatments of particle productions from the dissociation of the incident particle for each event. Hereafter, we call the latter one as particle productions from diffractive dissociation. This particle productions from diffractive dissociation depends on the diffractive mass.
The modeling of diffractive collisions have two characteristics and there are four types of the diffractive collision. Therefore, there are three characteristics in the diffractive collisions:
\begin{itemize}
    \item cross-section of each collision type,
    \item diffractive-mass spectrum, and
    \item particle productions from diffractive dissociation. 
\end{itemize}
Because the diffractive collisions are characterized by a low multiplicity and a high elasticity, air shower development shift deep in atmosphere~\cite{ulrich2011}. 
The collision types in the diffractive collisions also affect the UHECR observables as follows: If the first interaction of an air shower is a pSD or a DD collision, the projectile particle dissociates and produces several high-energy particles. However, if the first interaction is produced by a tSD or a CD collision, the projectile particle is intact; therefore, $\langle X_{\mathrm{max}}\rangle $ is predicted to become approximately one interaction length deep.  
The modeling of the diffractive collisions is also important. 
The diffractive collisions with smaller $M_X$ are characterized by using a smaller number and higher energy of produced particles, that results large $\langle X_{\mathrm{max}}\rangle $. Furthermore, there are differences in particle productions of the diffractive dissociation between the models, and these differences on the distribution of $M_X$ affect the uncertainty of the energy and the number of produced particles in the diffractive collisions. 

Recently, the cross-sections of the diffractive collisions were measured by TOTEM~\cite{TOTEM_diff_7tev,TOTEM_doublediff_7tev}, ATLAS~\cite{ATLAS_diff_7tev, ATLAS_crosssection_13tev,ATLAS_diff_ALFA}, CMS~\cite{CMS_diff_7tev}, and ALICE~\cite{ALICE_diff_7tev} at the Large Hadron Collider (LHC) in CERN. Since most of the particles are produced only in the very forward regions in the low diffractive-mass cases, measurements of the low diffractive-mass events with $M_X < 3.4$~GeV are limited. Therefore there are large uncertainties in the cross-section. 
Since the total and elastic cross-sections are precisely measured using Roman Pots by TOTEM~\cite{TOTEM_crosssection_7tev, TOTEM_crosssection_13tev} and ATLAS~\cite{ATLAS_optical_7tev}, the uncertainty in the inelastic cross-section is very small. 
However, the relative cross-sections between the collision types are equally important as the absolute value of the total and inelastic cross-sections. 
Moreover, particle productions from diffractive dissociation in the diffractive collisions are not well constrained by the collider experiments. 
Thus, these three quantities, such as cross-section of each collision type, diffractive-mass spectrum, and particle productions from diffractive dissociation, have large uncertainties.

\begin{figure}
    \centering
    \includegraphics[clip, width=0.9\columnwidth]{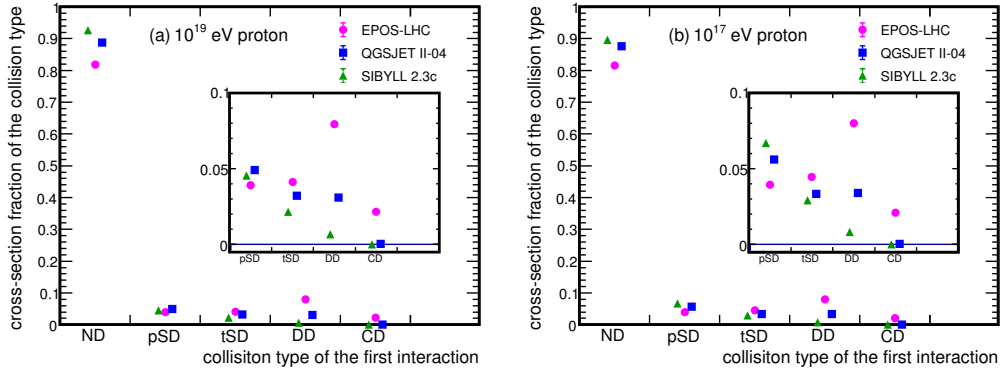}
    \caption{Cross-sectional fractions of the ND, pSD, tSD, DD, and CD collisions for (a) $10^{19}$~eV proton primaries and (b) $10^{17}$~eV proton primaries. 
    }
    \label{fig:FractionAtFirstInt}
\end{figure}

\begin{figure}
    \centering
    \includegraphics[clip, width=0.9\columnwidth]{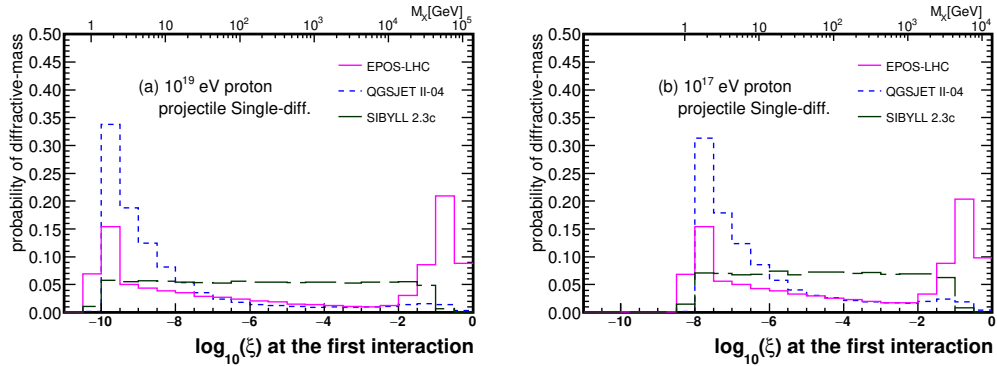}
    \caption{The probability of the diffractive mass (diffractive-mass spectra) of the pSD collisions for three hadronic interaction models: EPOS-LHC (magenta solid line), QGSJET~II-04 (blue dotted line), and SIBYLL~2.3c (green dashed line). The projectile particles are (a) $10^{19}$~eV protons and (b) $10^{17}$~eV protons, and the target particles are air nuclei at rest. Approximately 40,000 events are simulated for each case using CONEX~v6.40~\cite{conex}. 
    }
    \label{fig:DiffMassSpectrum}
\end{figure}

Large differences in the predictions of the diffractive collisions are found in the latest hadronic interaction models. 
Figure~\ref{fig:FractionAtFirstInt} displays the cross-sectional fraction in the proton-air collisions for $10^{19}$~eV and $10^{17}$~eV projectile protons. The difference of the cross-sectional fractions of the ND collisions between the hadronic interaction models is approximately 10\%, and SIBYLL~2.3c~\cite{SIBYLL1,SIBYLL2} and EPOS-LHC~\cite{EPOS} present the largest and smallest values, respectively. The difference of the cross-sectional fractions between models is relatively larger for the DD and CD collisions than that for the pSD and tSD collisions. 
Figure~\ref{fig:DiffMassSpectrum} displays the diffractive-mass spectra of the proton-air collisions for $10^{19}$~eV and $10^{17}$~eV projectile protons. Here, $\xi$ is defined as $\xi=M_{X}^2/s$, where $\sqrt{s}$ is the center-of-mass energy of the proton-air collision. The logarithm of $\xi$, $log_{10} (\xi)$ is used in Fig.~\ref{fig:DiffMassSpectrum} and following figures.
Large differences can be noted between SIBYLL~2.3c and the other models, particularly in the lowest diffractive-mass regions. QGSJET~II-04\cite{QGSJET} displays a strong peak in the low diffractive-mass region, whereas SIBYLL~2.3c does not present diffractive-mass dependencies and EPOS-LHC has a bimodal spectrum. Even though the latest hadronic interaction models are updated using the experimental results from the LHC, they do not reproduce the results of the measurements of the diffractive collisions from the LHC~\cite{Pierog_2019}. 


In the following sections, to understand the effect of these quantities in predicting the observables, the effects of the cross-sectional fraction, diffractive-mass spectrum, and particle productions from diffractive dissociation on the UHECR observables are discussed. The focus is on the large differences in the cross-sectional fractions of the models and in the diffractive-mass spectra of SIBYLL~2.3c and the other models.

\section{\label{sec:AirShowerSimulation}Air shower simulations}
\subsection{\label{sec:SimulationSetup}Simulation set-up}
Air shower events were simulated using CONEX~v6.40~\cite{conex} with three hadronic interaction models, EPOS-LHC, QGSJET~II-04, and SIBYLL~2.3c. UrQMD~\cite{UrQMD1,UrQMD2} was used for the collisions with incident particle energy less than 80~GeV. Primary protons of $10^{19}$ and $10^{17}$~eV with incident zenith angle 60$^{\circ}$ were assumed. The following three types of modifications of CONEX were applied: (1) the information on the collision type at the first interaction was additionally included in the output of each shower event. For the definition of the collision type, we used the collision information obtained from each hadronic interaction model, similar to in a previous study~\cite{ArbeletcheAirShower}. (2) To simulate the events used in Sec.~\ref{sec:XmaxFractionResampling}, the fractions of diffractive collisions were modified for the first interactions and following interactions of secondary particles. (3) For the study of the diffractive-mass dependence, as described in Sec.~\ref{sec:XmaxDiffMass}, the diffractive mass at the first interaction was calculated for each air shower from the momenta of the particles in the dissociation system. For these air shower events used in (3), the collision type at the first interaction was fixed to the pSD collision for the simplicity of the calculation of the diffractive mass.
For some models, an intact atmospheric nucleus is handled as nucleons after the collision even for the pSD collision. To separate these nucleons and the proton dissociation system, a rapidity threshold was used, which is set as 1.5 in the laboratory frame, and particles with rapidity larger than 1.5 were used for calculation of the diffractive mass. 
The number of produced events was 40,000 for each case. Although the calculation is effective for most of the events, less than 0.5~\% of the events have masses smaller than that of a proton, which is non-physical, owing to the miss-separation of the dissociation system in the calculations. These non-physical events were not used in the analysis. 

CONEX is sufficiently fast for simulating enough number of events because of employing the one-dimensional cascade equation. 
To confirm the predictions by CONEX, $\langle X_{\mathrm{max}}\rangle $ calculated by CONEX was compared to the result by a three-dimensional air shower simulation package, COSMOS~8.035~\cite{COSMOS}, for a $10^{17}$~eV proton primary case. To reduce the time-consuming calculation, a one-dimensional (1D) analytical calculation is conducted for the calculations of electromagnetic cascade showers in COSMOS. 
In this comparison, QGSJET~II-04 was used as a high-energy interaction model, and the zenith angle was 0$^{\circ}$. UrQMD was used for the low-energy interactions below 80~GeV in CONEX, whereas DPMJET~III~\cite{DPMJET} and PHITS~\cite{phits} were used in COSMOS for the interactions in 2-80~GeV and less than 2~GeV, respectively. In COSMOS, the number of electrons was calculated every 25~$\mathrm{g/cm^2}$, and $X_{\mathrm{max}}$ was calculated by fitting them to the Gaisser-Hillas function. 
The number of simulated air showers was 3,000 for each package. The result of $\langle X_{\mathrm{max}}\rangle $ was 683.4 $\pm$ 1.3~$\mathrm{g/cm^2}$ for CONEX and 680.9 $\pm$ 1.3~$\mathrm{g/cm^2}$ for COSMOS. The one-dimensional cascade equation in CONEX presents a reasonable agreement with the three-dimensional (3D) cascade simulation by COSMOS.

\subsection{\label{sec:Profiles}Longitudinal shower profiles and collision type}

\begin{figure}
    \centering
    \subfigure[Number of electrons and positrons]{%
        \includegraphics[clip, width=0.6\columnwidth]{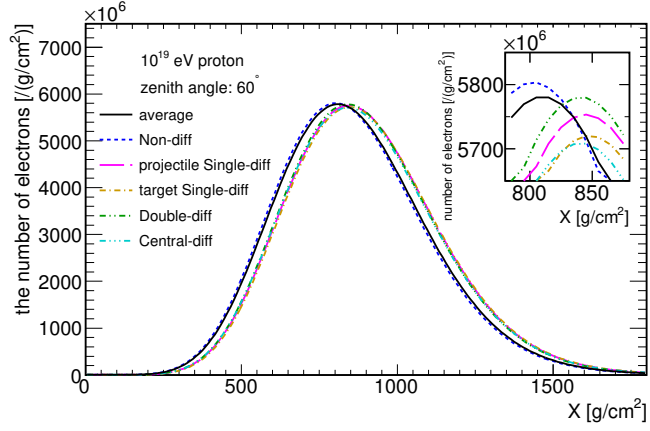}
    }
    \subfigure[Number of muons]{%
        \includegraphics[clip, width=0.6\columnwidth]{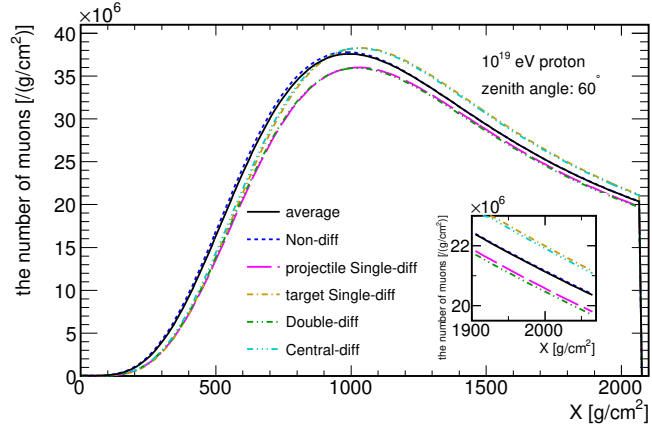}
    }
    \subfigure[Number of muon productions]{
        \includegraphics[clip, width=0.6\columnwidth]{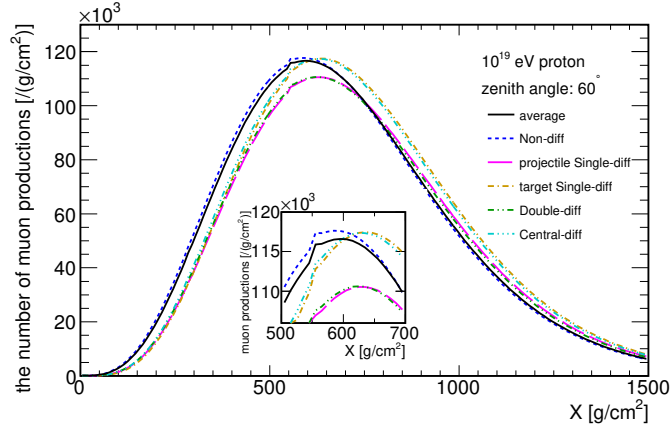}
    }
    \caption{The profiles of the average of (a) the number of electrons and positrons, (b) the number of muons, and (c) the number of muon productions in a $10^{19}$~eV proton shower with a zenith angle of 60$^{\circ}$ as a function of the atmospheric depth, $X$. Different lines represent the profiles categorized by the collision type at the first interaction.  EPOS-LHC is used for the hadronic interaction model. Electrons and positrons with energy more than 0.01~GeV and muons with energy more than 1~GeV are counted.} 
    \label{fig:MuonAndDepth}
\end{figure}

Before discussing about $ X_{\mathrm{max}} $ and $ X_{\mathrm{max}}^{\mu} $, the basic relationship between the diffractive collision type and the longitudinal air shower profile is discussed. To understand the relationship, the events are categorized by the collision type at the first interaction of the air shower, and the mean longitudinal shower profile is calculated for each category. The profiles of the electrons plus positrons (electrons hereafter unless mentioned) in $10^{19}$~eV showers with a zenith angle of 60$^{\circ}$ using EPOS-LHC are displayed in Fig.~\ref{fig:MuonAndDepth}~(a). The peak position of the diffractive collision categories is 40-50~$\mathrm{g/cm^{2}}$ deeper than that of the ND collision category, whereas the differences in categories of the diffractive collision are small. 

The shower profiles of the muons and muon productions using EPOS-LHC are presented in Figures~\ref{fig:MuonAndDepth}~(b) and (c), respectively. \footnote{Some structures can be found around the peak of muon production profiles. The hadronic interaction model was changed at 80~GeV in CONEX and it can create these structures.}
Similar to the electron profiles, the peak positions for the diffractive collision categories is 40-50~$\mathrm{g/cm^2}$ deeper than that by the ND. Additionally, the categories of the pSD and DD collision exhibit a smaller peak height than the other categories. For depth $X > 1100~\mathrm{g/cm^2}$, the number of muons of the tSD and CD collision categories is the largest among the five categories, whereas those of the pSD and DD collision categories are the smallest. This is because projectile cosmic rays are intact for the tSD and CD collisions while the projectile cosmic rays dissociate and low multiplicity collision are expected for the pSD and DD collisions. 
Moreover, the difference in the number of muons of the categories depends on the depth, and it decreases in a deeper atmosphere. 
Diffractive collisions are characterized by the high elasticity, and that make the shower developments deep. For the pSD and DD collision categories, the peak height is small while the shower developments is deeper than the ND collision category. 
Therefore the differences between the ND and pSD collision categories in a deeper atmosphere is small. 
At $X=2000\,\mathrm{g/cm^2}$, the differences between the ND collision category and the diffractive collision categories are less than 3\,\%. Considering small fractions of diffractive collisions, the effects of diffractive collisions at the first interaction is expected to be less than 0.3\,\%. It is noted that diffractive collisions of interactions of secondary particles is not considered while they also affect the number of muons. 

These mean longitudinal profiles are directly related to the observable variables in the air shower experiments.
The peak depth of the electron profile is defined by $ X_{\mathrm{max}} $, and on the peak depth of the muon production point is defined by $ X_{\mathrm{max}}^{\mu} $. 
It is found that the diffractive collisions at the first interaction make $ X_{\mathrm{max}} $ and $ X_{\mathrm{max}}^{\mu} $ larger. 
Another observable is the number of muons on the ground, $N_{\mu}$. 
Since the effect of diffractive collisions on the number of muons at $X=2000\,\mathrm{g/cm^2}$, which is around the ground for inclined showers, is expected to be small, we only focus on $ X_{\mathrm{max}} $ and $ X_{\mathrm{max}}^{\mu} $. 
The quantitative discussions of the model dependencies and the effects to $\langle X_{\mathrm{max}}\rangle $ and $\langle X_{\mathrm{max}}^{\mu}\rangle $ are discussed in Sec.~\ref{sec:EffectOnXmax} with focusing on the detail characteristics of diffractive collisions, such as cross-section of each collision type, the diffractive-mass spectrum, and particle productions from diffractive dissociation. The effects on the fluctuation of $X_{\mathrm{max}}$ are discussed in Sec.~\ref{sec:SigmaXmax}.

\section{\label{sec:EffectOnXmax}Effect of diffractive collisions on $\langle X_{\mathrm{max}}\rangle $ and $\langle X_{\mathrm{max}}^{\mu}\rangle $}

\subsection{\label{sec:XmaxFeature} Collision type and its effect to $\langle X_{\mathrm{max}}\rangle $, $\langle X_{\mathrm{max}}^{\mu}\rangle $ }

\begin{figure}
    \centering
    \includegraphics[clip, width=0.9\columnwidth]{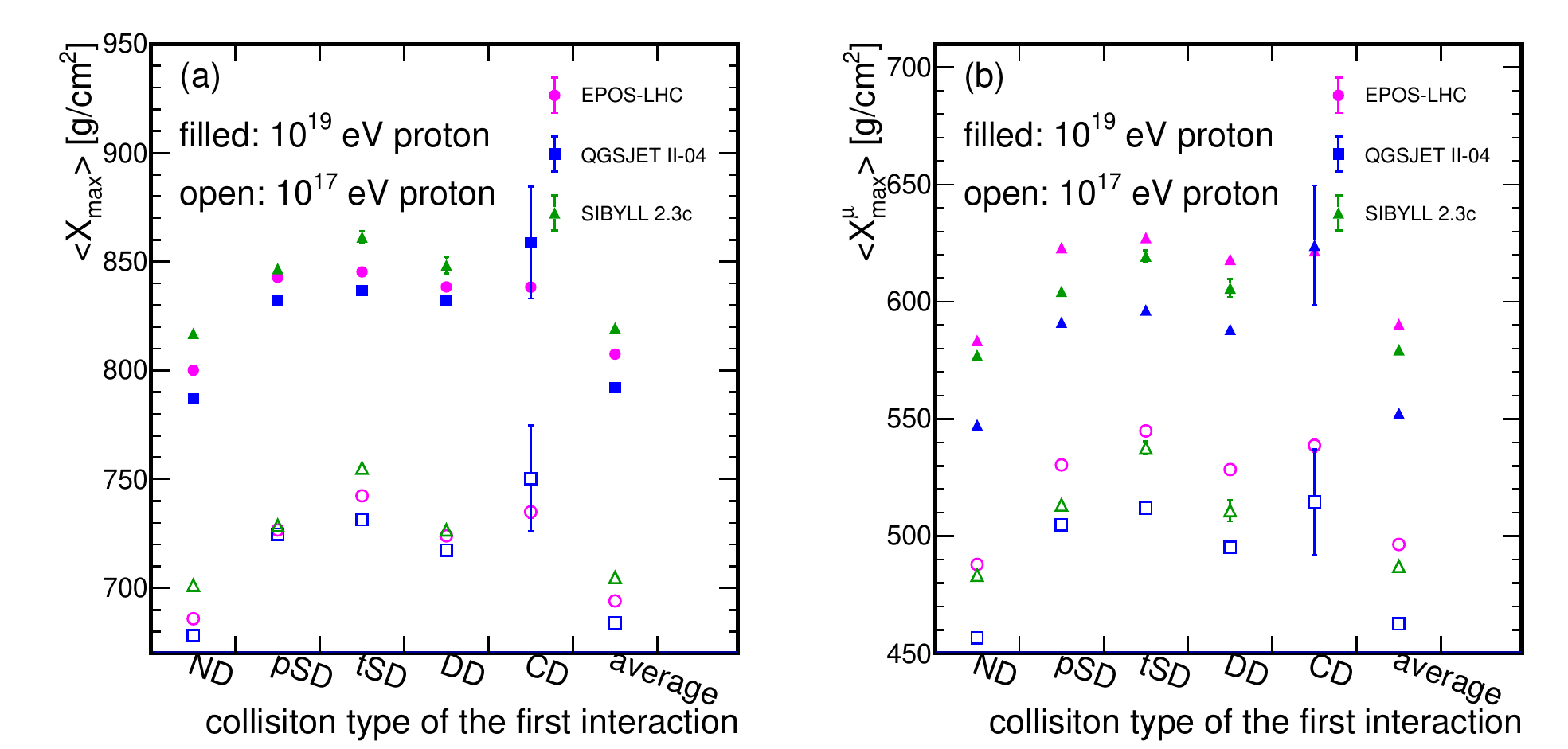}
    \caption{(a) $\langle X_{\mathrm{max}}\rangle $ and (b) $\langle X_{\mathrm{max}}^{\mu}\rangle $ categorized by the collision type at the first interaction and the average. The energy of the primary proton is $10^{17}$~eV (open symbols) and $10^{19}$~eV (filled symbols), and EPOS-LHC (magenta), QGSJET~II-04 (blue), and SIBYLL~2.3c (green) are used for the interaction model.}
    \label{fig:XmaxWithCategory}
\end{figure}

$\langle X_{\mathrm{max}}\rangle $ and $\langle X_{\mathrm{max}}^{\mu}\rangle $ are calculated for each event category at the first interaction,
as displayed in Figures~\ref{fig:XmaxWithCategory}~(a) and (b), respectively. We can find differences between predictions of collision type categories even in one model, and the prediction of the tSD collision category shows the largest among the categories. This feature means that the collision type at the first interaction affect $\langle X_{\mathrm{max}}\rangle $ and $\langle X_{\mathrm{max}}^{\mu}\rangle $ as follows; if the cross-sectional fraction of the tSD collision become larger, 
the average of the $\langle X_{\mathrm{max}}\rangle $ prediction become larger. 
Moreover, even in the same category, there are differences in the predictions of the models. 
In this section, we discuss the effects of cross-sectional fraction between collision types and the effects of the modeling of diffractive collisions.

In Figures~\ref{fig:XmaxWithCategory}~(a) and (b), SIBYLL~2.3c displays large differences of $\langle X_{\mathrm{max}}\rangle $ and $\langle X_{\mathrm{max}}^{\mu}\rangle $ between the pSD and tSD collision categories, which is 20-30~$\mathrm{g/cm^2}$. Moreover, the size of the difference between the pSD and tSD collision categories is as large as the difference between the ND and pSD collisions categories.
For EPOS-LHC, the differences between the pSD, tSD, and DD collision categories are much smaller than those between the pSD and ND collision categories. These features suggest that the cross-sectional fraction of the diffractive collisions to the ND collisions is more important than that of the pSD and tSD collision for the EPOS-LHC cases. In comparison, the cross-sectional fractions of the pSD and tSD collision categories may affect $\langle X_{\mathrm{max}}\rangle $ and $\langle X_{\mathrm{max}}^{\mu}\rangle $ in the SIBYLL~2.3c case. Details of the effects of the cross-sectional fractions are discussed in the following section. 

Focusing on the values in each category of the collision type, the size of the difference between SIBYLL~2.3c and the other models vary with the collision categories, while the difference between QGSJET~II-04 and EPOS-LHC is found to be mostly independent of the collision category. 
For example, the differences in $\langle X_{\mathrm{max}}^{\mu}\rangle $ between SIBYLL~2.3c and EPOS-LHC for the pSD and the DD collision categories are larger than these for the ND and the tSD collision categories. The projectile protons (cosmic rays) dissociate only for the pSD and the DD collisions, therefore, the diffractive mass spectrum affect the shower development only in the pSD and the DD case. 
Thus, this can be related with a large difference of the diffractive mass spectrum between SIBYLL~2.3c and the other models. It is important to examine the effects of the modeling of the diffractive collisions, including the effect of the diffractive mass spectrum on $\langle X_{\mathrm{max}}\rangle $ and $\langle X_{\mathrm{max}}^{\mu}\rangle $.

\subsection{\label{sec:XmaxFraction}Effect of cross-sectional fractions of the collision types on $\langle X_{\mathrm{max}}\rangle $ and $\langle X_{\mathrm{max}}^{\mu}\rangle $}

The contribution of the cross-sectional fractions is examined from two points of view.
One is the effect of the first interaction in the air showers focusing on all the cross-sectional fractions. The other is the effect including the interactions of secondary particles over the whole air shower, focusing on one cross-sectional fraction that displays the largest effect at the first interaction.

\subsubsection{\label{sec:XmaxFractionFirstint}Effect of first interaction}

\begin{figure}
    \centering
        \includegraphics[clip, width=0.5\columnwidth]{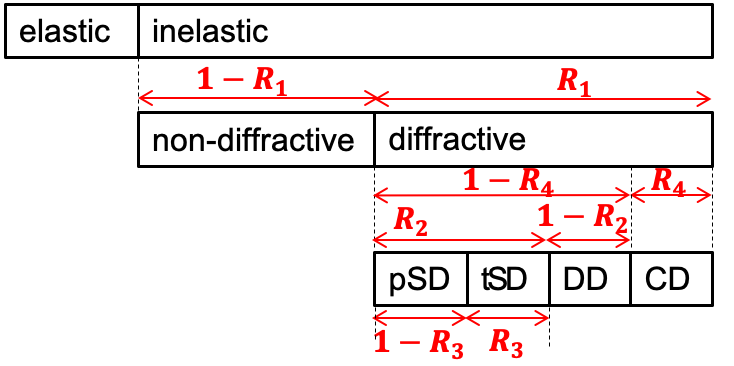}
    \caption{Schematic view of four ratios $R_1$, $R_2$, $R_3$, and $R_4$}
    \label{fig:CharacteristicsAndAnalyses}
\end{figure}

The effects of the cross-sectional fractions at the first interaction are studied using $\langle X_{\mathrm{max}}\rangle $ and $\langle X_{\mathrm{max}}^{\mu}\rangle $, as introduced in Fig.~\ref{fig:XmaxWithCategory}. The average of $\langle X_{\mathrm{max}} \rangle $, $\langle X_{\mathrm{max}}^{\mathrm{average}}\rangle$, can be calculated from the $\langle X_{\mathrm{max}} \rangle$ of each category shown in Fig.~\ref{fig:XmaxWithCategory} considering the cross-section fractions of each category as follows;
\begin{equation}
\langle X_{\mathrm{max}}^{\mathrm{average}}\rangle = \sum_{i} f^{i} \langle X_{\mathrm{max}}^{i}\rangle ,
\label{eq:Xmax_sum}
\end{equation}
where $f^{i}$ is the cross-sectional fraction of the $i$-th collision type. 
$\langle X_{\mathrm{max}}^{\mu, \mathrm{average}}\rangle $ can be calculated similarly as above.
We can easily estimate the effects of the cross-sectional fraction on $\langle X_{\mathrm{max}}^{\mathrm{average}}\rangle $ and $\langle X_{\mathrm{max}}^{\mu,\,\mathrm{average}}\rangle $ by changing $f^{i}$ artificially in equation~\ref{eq:Xmax_sum}.
Hereafter, the expected value of $\langle X_{\mathrm{max}}^{\mathrm{average}}\rangle $ after the artificial modification in $f^{i}$ is referred as $\langle X_{\mathrm{max}}^{\mathrm{modified}}\rangle $.
Using this method, the effects of the cross-sectional fractions can be studied without repeating the time-consuming MC simulations. 
For example, if the cross-sectional fraction of the tSD collisions are increased while the fraction of the ND collisions is reduced, $\langle X_{\mathrm{max}}^{\mathrm{modified}}\rangle$ become larger. 
Instead of directly changing $f^{i}$'s, the four ratios of the cross sections, $R_{1}$, $R_{2}$, $R_{3}$, and $R_{4}$, is introduced. As illustrated in Fig.\ref{fig:CharacteristicsAndAnalyses}, the ratios are defined between 0 and 1. 
$R_{1}$ is the ratio of all the diffractive collisions to the inelastic collisions. $R_{2}$ is the ratio of the SD  (pSD + tSD) collisions to the sum of the SD and DD collisions. $R_{3}$ is the ratio of the tSD collisions to all the SD collisions. $R_{4}$ is the ratio of the CD collisions to all the diffractive collisions. The cross-sectional fractions, $f^i$'s, can be calculated from these four ratios as, 

\begin{align}
\begin{split}
f^{ND} &= 1-R_{1},\\
f^{pSD} &= R_{1} (1-R_{4}) R_{2} (1-R_{3}),\\
f^{tSD} &= R_{1} (1-R_{4}) R_{2} R_{3},\\
f^{DD} &= R_{1} (1-R_{4}) (1-R_{2} ), and\\
f^{CD} &= R_{1} R_{4}.
\end{split}
\label{eq:fraction}
\end{align}
$\langle X_{\mathrm{max}}^{\mathrm{modified}}\rangle $ and $\langle X_{\mathrm{max}}^{\mu,\mathrm{modified}}\rangle $ are calculated using the cross-sectional fractions and Eq.~\ref{eq:Xmax_sum} by changing the ratios.

\begin{figure}
    \centering
    \subfigure[$R_1$ dependence of $\langle X_{\mathrm{max}}^{\mathrm{modified}}\rangle $]{%
        \includegraphics[clip, width=0.4\columnwidth]{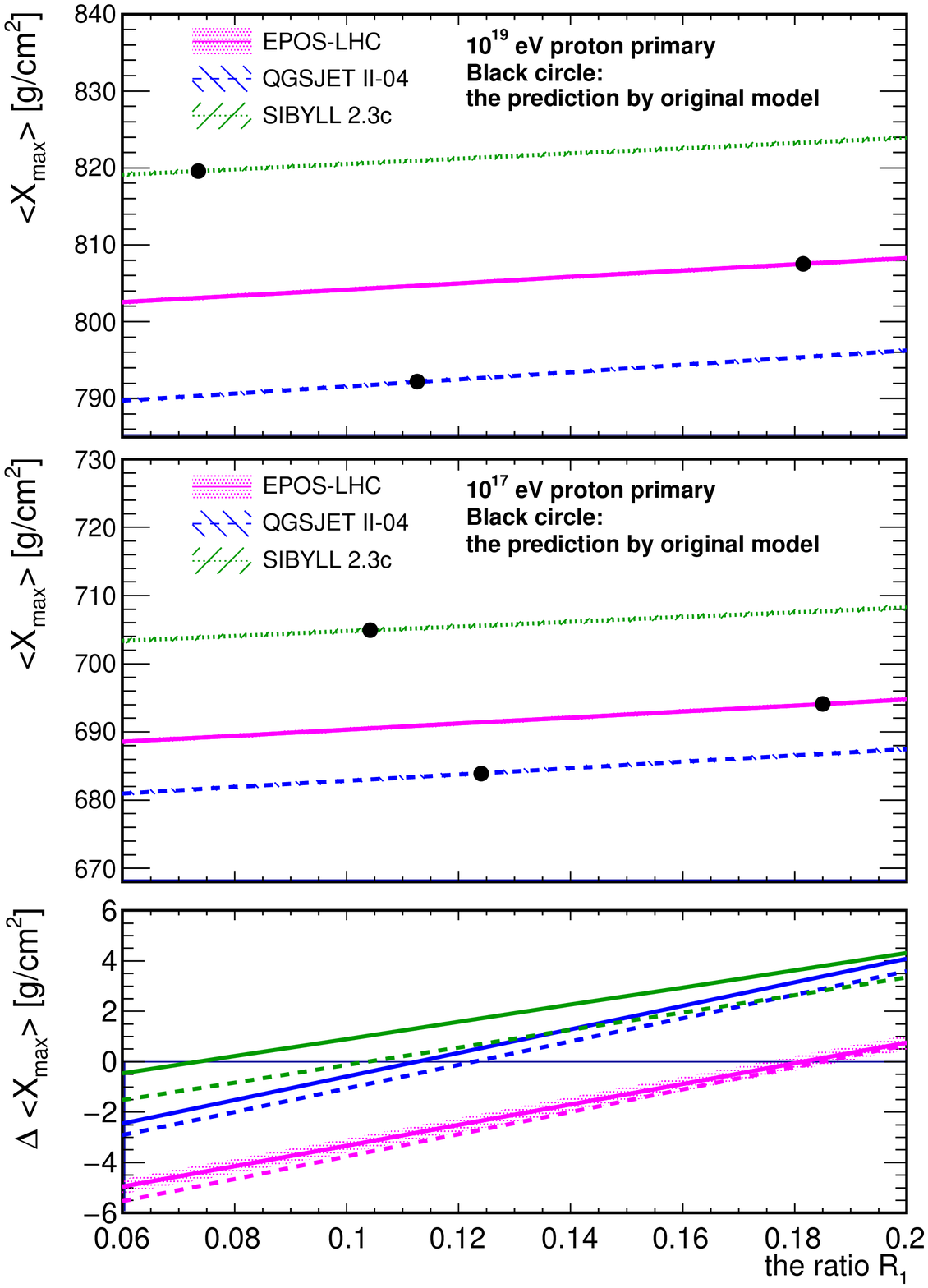}
    }%
    \subfigure[$R_2$ dependence of $\langle X_{\mathrm{max}}^{\mathrm{modified}}\rangle $]{%
        \includegraphics[clip, width=0.4\columnwidth]{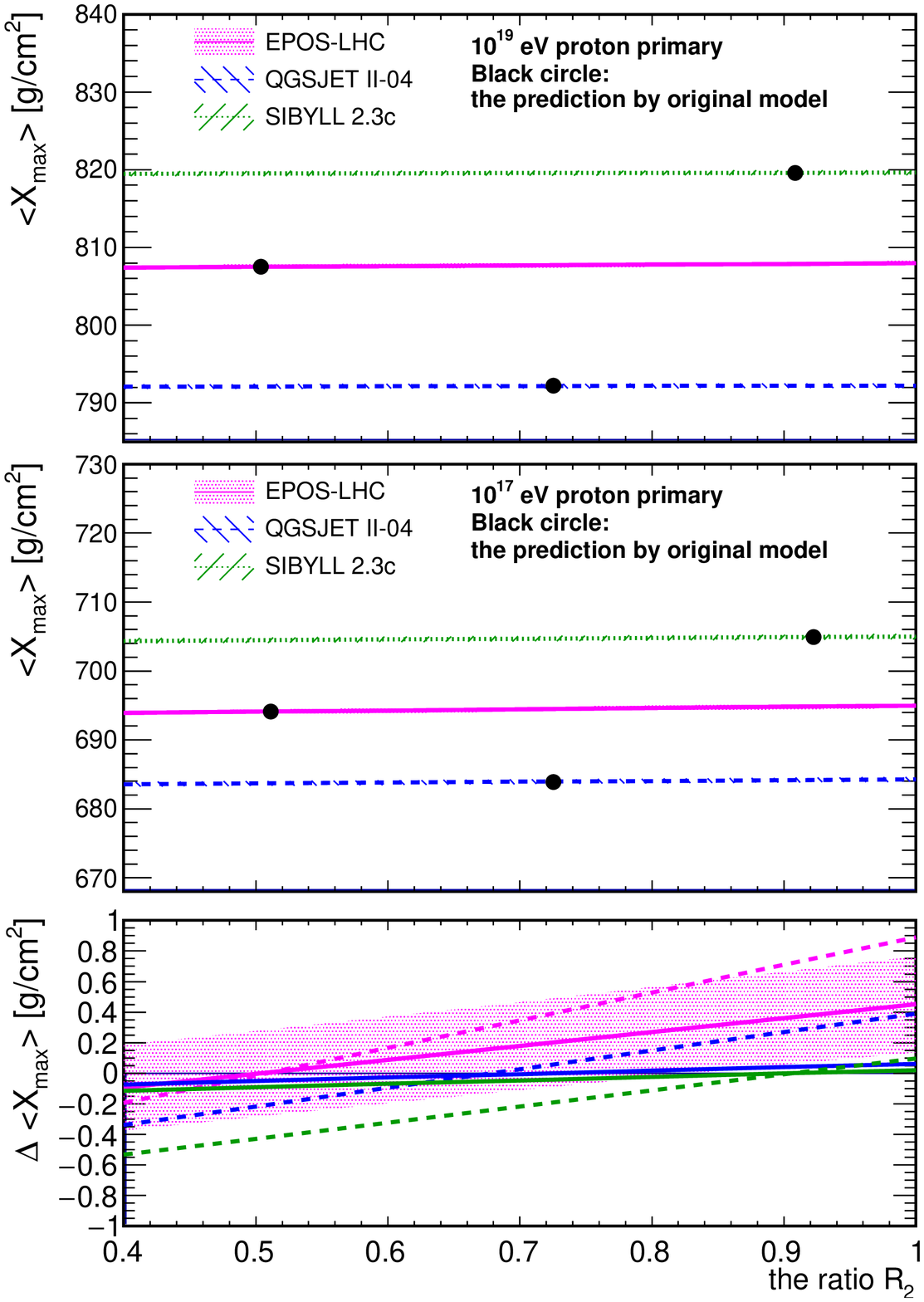}
    }%
    \\
    \subfigure[$R_3$ dependence of $\langle X_{\mathrm{max}}^{\mathrm{modified}}\rangle $]{
        \includegraphics[clip, width=0.4\columnwidth]{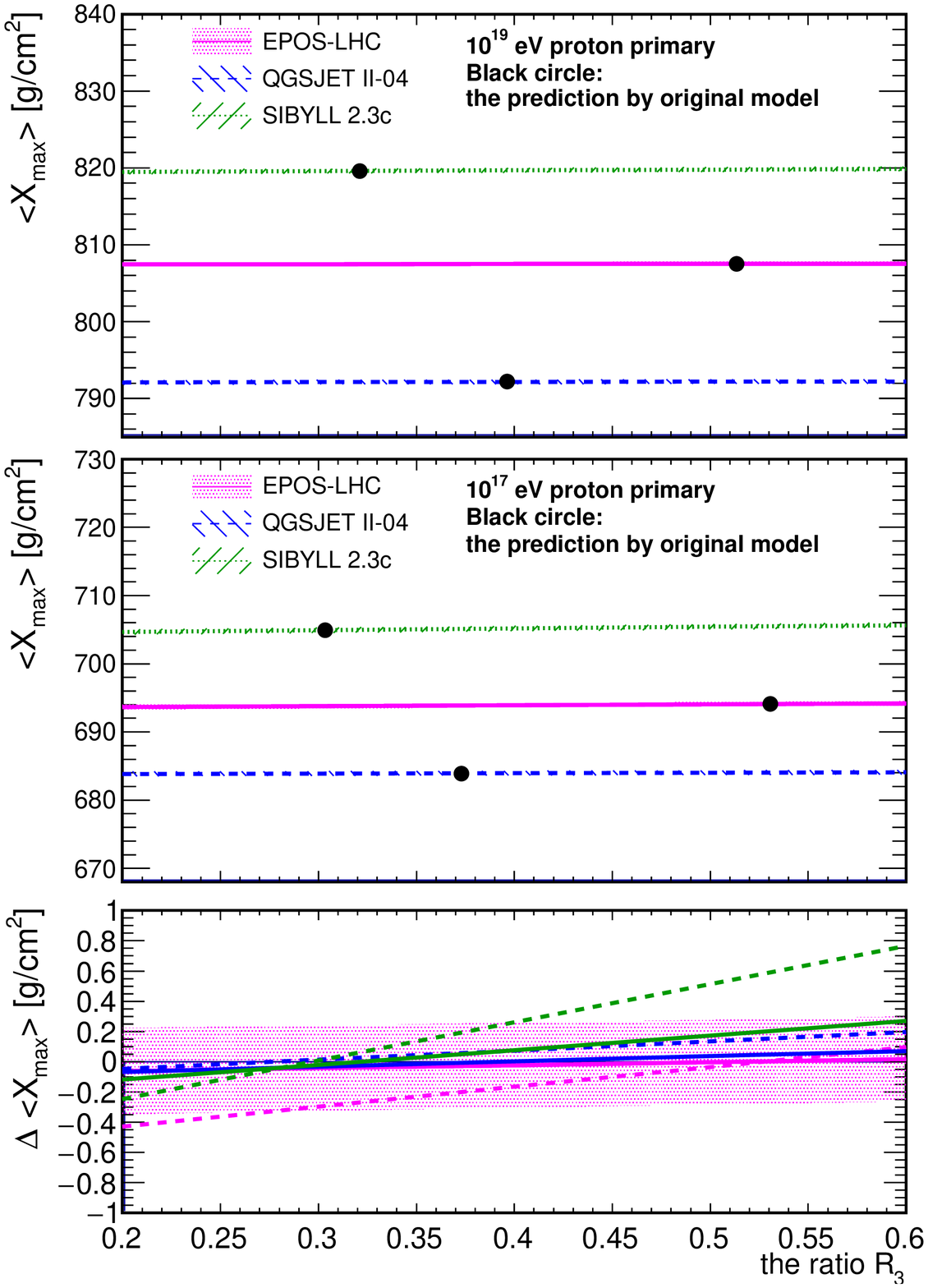}
    }
    \caption{The $\langle X_{\mathrm{max}}^{\mathrm{modified}}\rangle $ dependency for (a) the ratio $R_1$, (b) $R_2$, and (c) $R_3$. Hatched regions display the statistical errors. The upper and middle panels present the $\langle X_{\mathrm{max}}^{\mathrm{modified}}\rangle $ for the $10^{19}$~eV and $10^{17}$~eV proton primaries, respectively. Black circles denote the predictions with the original $R$ of the model.  The bottom panel displays $\Delta \langle X_{\mathrm{max}}\rangle $.
    Magenta solid lines denote the results of EPOS-LHC for $10^{19}$~eV ($10^{17}$~eV), the blue dashed lines represent the results of QGSJET~II-04 for $10^{19}$~eV ($10^{17}$~eV), and the green dotted lines display the results of SIBYLL~2.3c for $10^{19}$~eV ($10^{17}$~eV). In the bottom plots, the solid (dashed) lines present the results for $10^{19}$~eV ($10^{17}$~eV),  and the statistical errors of EPOS-LHC for $10^{19}$~eV only are displayed.}
    \label{fig:ModifyFraction}
\end{figure}

\begin{figure}
    \centering
    \subfigure[$R_1$ dependence of $\langle X_{\mathrm{max}}^{\mu, \mathrm{modified}}\rangle $]{%
        \includegraphics[clip, width=0.4\columnwidth]{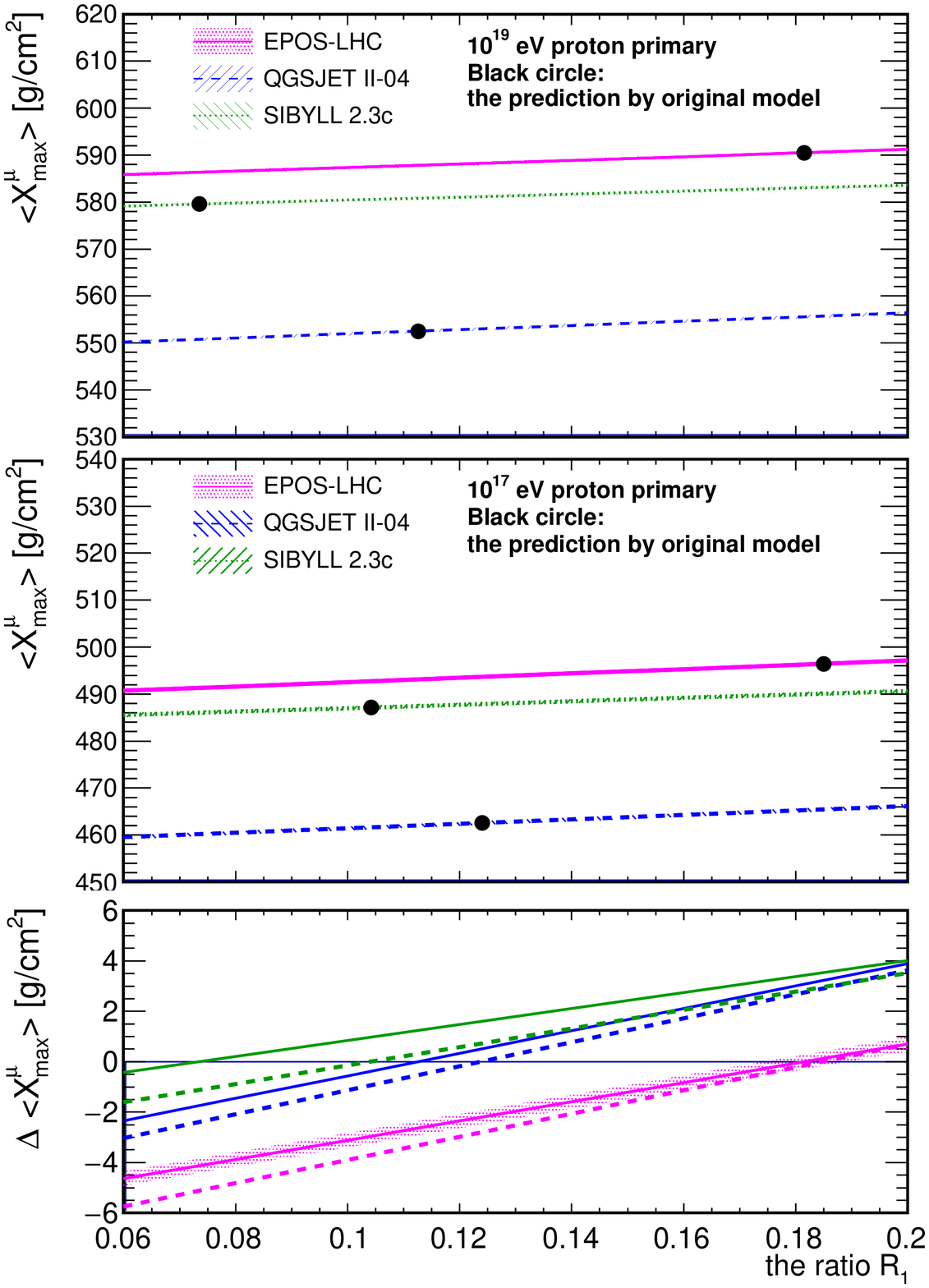}
    }%
    \subfigure[$R_2$ dependence of $\langle X_{\mathrm{max}}^{\mu, \mathrm{modified}}\rangle $]{%
        \includegraphics[clip, width=0.4\columnwidth]{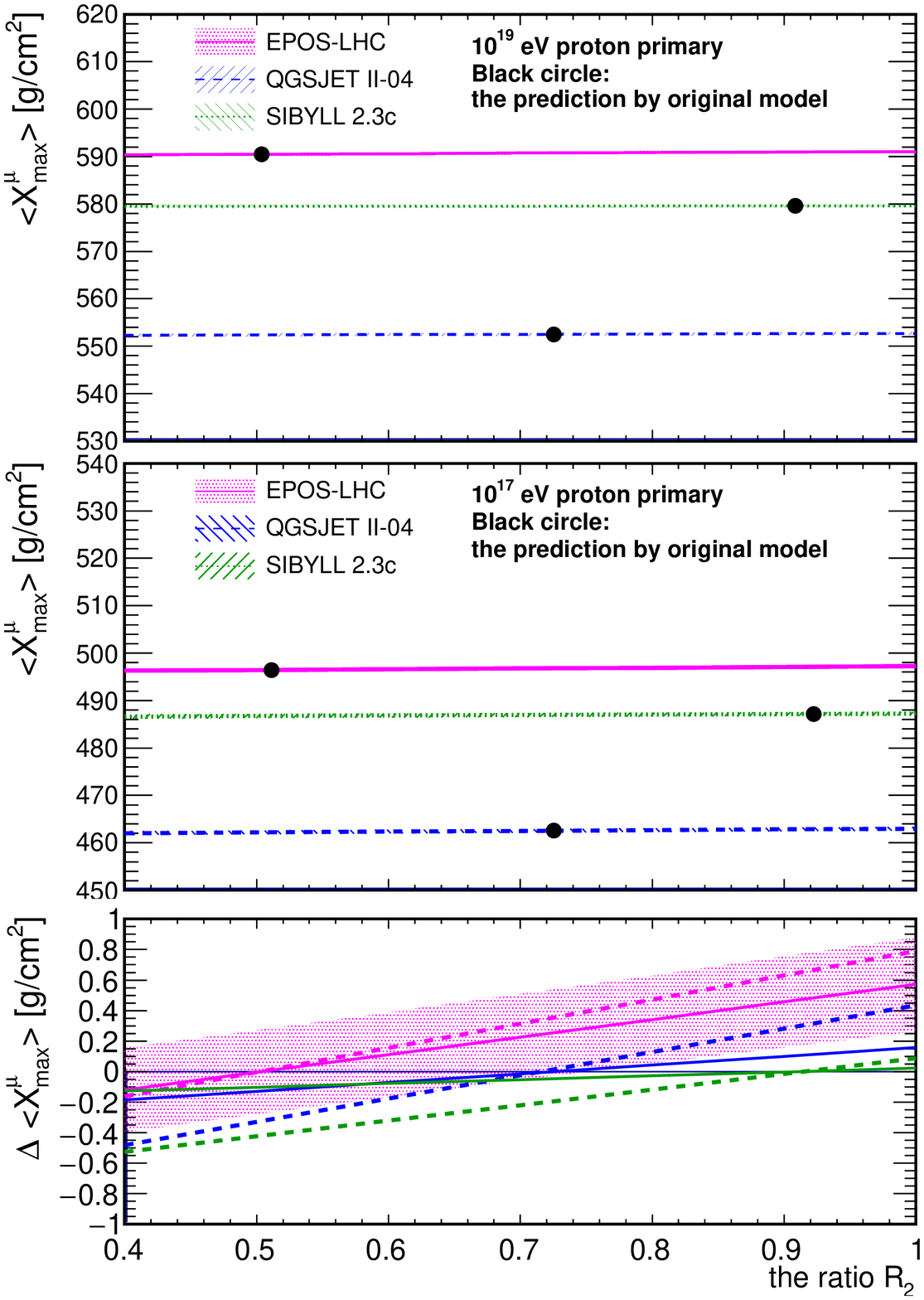}
    }%
    \\
    \subfigure[$R_3$ dependence of $\langle X_{\mathrm{max}}^{\mu, \mathrm{modified}}\rangle $]{
        \includegraphics[clip, width=0.4\columnwidth]{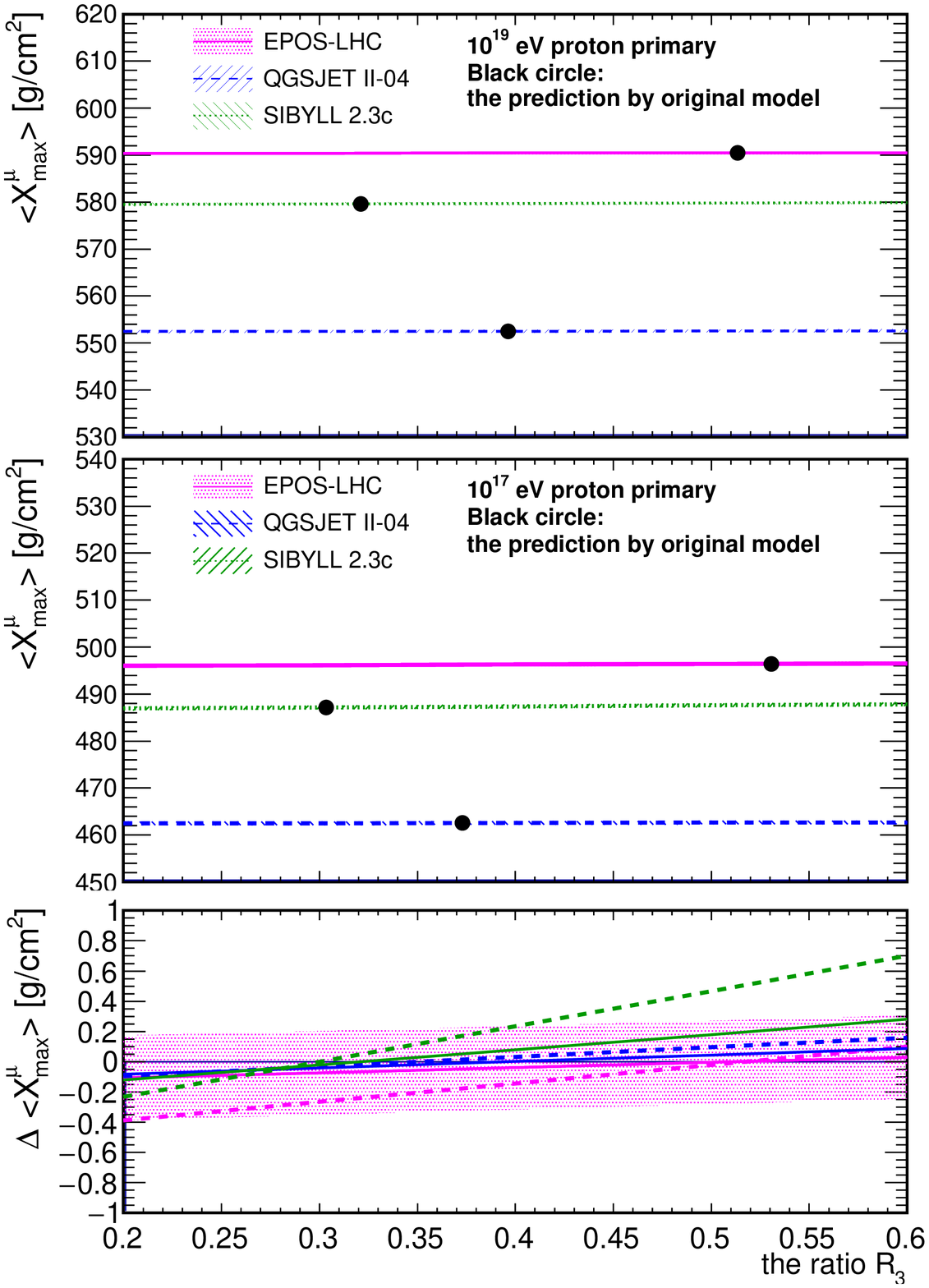}
    }
    \caption{The $\langle X_{\mathrm{max}}^{\mu, \mathrm{modified}}\rangle $ dependency on the ratio of $R_1$, $R_2$, and $R_3$. 
    }
    \label{fig:ModifyFractionMuon}
\end{figure}

The results of changing the cross-sectional fractions varying the ratios $R_{1}$, $R_{2}$ and $R_{3}$ are displayed in Figures~\ref{fig:ModifyFraction} and \ref{fig:ModifyFractionMuon}. Upper panels display $\langle X_{\mathrm{max}}^{\mathrm{modified}}\rangle $ and $\langle X_{\mathrm{max}}^{\mu , \mathrm{modified}}\rangle $ (lines) and the original predictions $\langle X_{\mathrm{max}}^{\mathrm{original}}\rangle $ and $\langle X_{\mathrm{max}}^{\mu,\mathrm{original}}\rangle $ (black circles) by three interaction models for $10^{19}$~eV proton primaries. 
The middle plots are the same as the upper plots except for the $10^{17}$~eV proton primaries. 
Magenta (solid and dash-dotted), blue (dashed and dash-two-dotted), and green (dotted and dash-three-dotted) lines display the results of EPOS-LHC, QGSJET~II-04, and SIBYLL~2.3c models, respectively.
The error bars of these plots display the statistical errors of the MC simulations. 
The bottom plots present the differences in $\langle X_{\mathrm{max}}^{\mathrm{modified}}\rangle $ and $\langle X_{\mathrm{max}}^{\mu , \mathrm{modified}}\rangle $ from the original predictions, where $\Delta \langle X_{\mathrm{max}}\rangle $ is defined as
\begin{equation}
\Delta \langle X_{\mathrm{max}}\rangle = \langle X_{\mathrm{max}}^{\mathrm{modified}} \rangle - \langle X_{\mathrm{max}}^{\mathrm{original}}\rangle .
\label{eq:DelataMeanXmax}
\end{equation}
In the bottom plots, the error bars only for the $10^{19}$~eV EPOS-LHC case are drawn for visibility.
The calculation for $R_4$ is not performed, because SIBYLL~2.3c does not include CD collisions.

Based on the upper plot of Fig.~\ref{fig:ModifyFraction}~(a), the original $R_{1}$, which is equal to cross-section fraction of diffractive collisions in inelastic collisions, ranges from 0.074 (SIBYLL~2.3c) to 0.182 (EPOS-LHC), which is regarded as the model uncertainty of $R_{1}$ in this analysis.
When $R_1$ is changed in this range,
$\langle X_{\mathrm{max}}\rangle $ changes by 3.7~$\mathrm{g/cm^2}$ for SIBYLL~2.3c and 5.0~$\mathrm{g/cm^2}$ for QGSJET~II-04, and $\langle X_{\mathrm{max}}^{\mu}\rangle $ changes by 3.4~$\mathrm{g/cm^2}$ for SIBYLL~2.3c and 4.8~$\mathrm{g/cm^2}$ for QGSJET~II-04 for $10^{19}$~eV. 
However, the shifts for the change of the ratios of $R_{2}$ and $R_{3}$ within the model uncertainties are smaller than 0.4~$\mathrm{g/cm^2}$, which is comparable to the statistical errors. These results signify that the cross-section ratio of the diffractive collisions to the inelastic collisions $R_{1}$ has a dominant effect on $\langle X_{\mathrm{max}}\rangle $ and $\langle X_{\mathrm{max}}^{\mu}\rangle $. 
On the other hand, the cross-section ratios of the diffractive collision types {\it i. e.} $R_2$ and $R_3$, are not sensitive to $\langle X_{\mathrm{max}}\rangle $ and $\langle X_{\mathrm{max}}^{\mu}\rangle $. 
The ratio dependences of $10^{17}$~eV for $R_1$ are similar to those of $10^{19}$~eV.

\subsubsection{\label{sec:XmaxFractionResampling}Effect of interactions over a whole air shower}

The effects of the interactions over a whole air shower including interactions of secondary particles on $\langle X_{\mathrm{max}}\rangle $ and $\langle X_{\mathrm{max}}^{\mu}\rangle $ are discussed in this section, focusing on the cross-sectional ratio of the diffractive collisions to inelastic collisions $R_{1}$. It exhibits the largest effect at the first interaction, as presented in Sec.\ref{sec:XmaxFractionFirstint}. 
In this study, the ratio for collisions with projectile energy larger than $10^{15}$~eV was modified. This threshold energy is chosen in a previous study using CONEX~\cite{ulrich2011}. 

\begin{figure}
    \centering
    \includegraphics[clip, width=0.45\columnwidth]{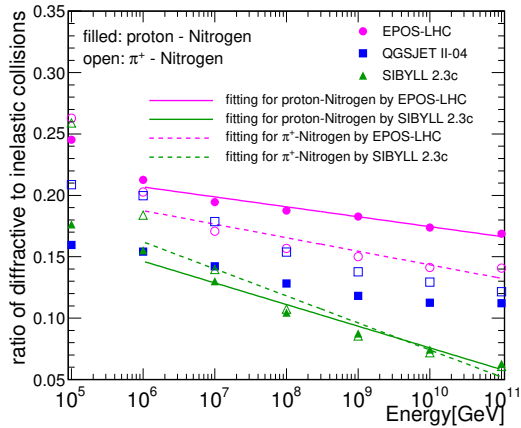}
    \caption{Energy dependencies of the cross-section ratio, $R_1$. The ratios of the proton-Nitrogen collisions (filled markers) and $\pi^{+}$-Nitrogen collisions (open markers) for three hadronic interaction models using CRMC v1.7\cite{CRMC} are displayed. Straight lines are the fitted results of the cross-section ratios with $a\log_{10}{E} + b$. The magenta solid (dashed) line is the fitted result for the proton ($\pi^{+}$)-Nitrogen collisions by EPOS-LHC, and the green solid (dashed) line is the fitted result for proton ($\pi^{+}$) Nitrogen collisions by SIBYLL~2.3c.}
    \label{fig:FractionFitting}
\end{figure}

Before the modification, we defined the reference $R_1$ ratios, $R_1^{\mathrm{Ref}}$, which are used for the reference of the ratio in the modification procedure, as follows; $R_1^{\mathrm{Ref}}$ as a function of energy are estimated from the simulated ratios of EPOS-LHC and SIBYLL~2.3c, as displayed in Fig.~\ref{fig:FractionFitting} by fitting with a function defined as $a\log_{10}{E} + b$, where $a$ and $b$ are free parameters. The magenta (green) solid and the dashed lines are the fitting results of the ratio for proton-Nitrogen and $\pi^{+}$-Nitrogen collisions simulated by EPOS-LHC (SIBYLL 2.3c), respectively. The fitting results of the proton-Nitrogen ($\pi^{+}$-Nitrogen) collisions are used as $R_1^{\mathrm{Ref}}$ for baryon-air nucleus (meson-air nucleus) collisions at a given energy in an air shower. Hereby, the modification with $R_1^{\mathrm{Ref}}$ taken from EPOS-LHC and SIBYLL~2.3c results is called as the EPOS-based $R_1^{\mathrm{Ref}}$ case and the SIBYLL-based $R_1^{\mathrm{Ref}}$ case, respectively.
Then, cross-sectional fractions of collision types are modified for each collisions in an air shower. The procedure of this modification is described below; %
\begin{enumerate}
\item For every collision with a projectile energy larger than $10^{15}$~eV in an air shower, a collision type is selected before the collision generation by following the cross-sectional fraction, $f_i$, using $R_1^{\mathrm{Ref}}$. 
\item 
A collision is generated according to each model, and the collision type of the produced collision is compared to the one selected in (1). If the two types are not the same, another collision is generated repeatedly for this collision until those two types become the same. If the two types are the same, the produced collision is accepted and the simulation of the air shower is proceeded.
\end{enumerate}

Results of $\langle X_{\mathrm{max}}\rangle $ with this modification are summarized in Tab.~\ref{tab:resampling}. 
The difference between the value of EPOS-LHC and the one obtained with the modification using EPOS-based $R_1^{\mathrm{Ref}}$ accounts for the systematic uncertainty of the method,
and the size of the difference is 2.0~$\mathrm{g/cm^2}$. The difference for the SIBYLL~2.3c case is 0.9~$\mathrm{g/cm^2}$.  
This systematic uncertainty is due to the difference between the original prediction of $R_1$ and $R_1^{\mathrm{Ref}}$, which is defined by fitting. We discuss the impact of the change in $R_1$ from the original prediction to the reference one. 
The effect of the ratio $R_1$ is 8.9$\pm$0.4~$\mathrm{g/cm^2}$ at the maximum for the EPOS-LHC-based case and 4.2$\pm$0.4~$\mathrm{g/cm^2}$ at the maximum for the SIBYLL~2.3c-based case, and they are 1-2 times as large effects in comparison with the value estimated from the cross-section ratio at the first interaction, 3.7-5.0~$\mathrm{g/cm^2}$. From the comparison between the estimation at the first interactions and the estimation with modification for whole air shower, it is suggested that the effects of the ratios at the first interaction and the interactions of secondary particles are equally important for the $\langle X_{\mathrm{max}}\rangle $ predictions. 

There is another important point. The maximum differences in the $\langle X_{\mathrm{max}}\rangle $ predictions between the models with this modification are 31.9$\pm$0.5~$\mathrm{g/cm^2}$ for the EPOS-LHC-based case and 32.0$\pm$0.4~$\mathrm{g/cm^2}$ for the SIBYLL~2.3c-based case, as displayed in the bottom row of Tab.~\ref{tab:resampling}. Moreover, they are larger than the difference without any modification: 27.4$\pm$0.4~$\mathrm{g/cm^2}$. 
This means that even if the cross-section ratio, $R_1$, is fixed to the same value in three models, the difference in the prediction $\langle X_{\mathrm{max}}\rangle $ between the models become large. 
This result may be caused by the other sources of the differences in $\langle X_{\mathrm{max}}\rangle$ predictions; in the original models, several sources contribute the difference of $\langle X_{\mathrm{max}}\rangle $ predictions and the differences of $R_1$ in models reduce the difference of $\langle X_{\mathrm{max}}\rangle $ predictions by chance. 
The cross-section ratios affect the mean value of $\langle X_{\mathrm{max}}\rangle $ predictions; however, they cannot solve the differences in the $\langle X_{\mathrm{max}}\rangle $ predictions between models. 

$\langle X_{\mathrm{max}}^{\mu}\rangle $ can be similarly discussed based on Tab. 2. 
The effects of the cross-section ratio, $R_1$, are 9.4$\pm$0.4~$\mathrm{g/cm^2}$ for the EPOS-LHC-based case and 4.4$\pm$0.4~$\mathrm{g/cm^2}$ for the SIBYLL~2.3c-based case at maximum.
The differences in the $\langle X_{\mathrm{max}}^{\mu}\rangle $ predictions between the models, with this modification, are smaller by 5.2$\pm$0.6~$\mathrm{g/cm^2}$ and 0.4$\pm$0.6~$\mathrm{g/cm^2}$ than those of the original predictions for the EPOS-LHC-based case and the SIBYLL~2.3c-based case, respectively. 
The cross-section ratios affect the mean value of the $\langle X_{\mathrm{max}}^{\mu}\rangle $ predictions; however, they cannot solve the differences in the $\langle X_{\mathrm{max}}^{\mu}\rangle $ predictions of the models.

\begin{table}[!h]
    \centering
    \caption{Results of $\langle X_{\mathrm{max}}\rangle $ with cross-section ratio modifications over the whole air shower. The reference ratio, $R_1^{\mathrm{Ref}}$, for the modification is set by fitting the ratio of the predictions by EPOS-LHC or SIBYLL~2.3c with energy above $10^{15}$~eV.}
    \label{tab:resampling}
    \footnotesize
    \begin{tabular}{l||c|cc|cc}
    \hline
    &\multicolumn{4}{c}{$\langle X_{\mathrm{max}}\rangle $ [g/cm$^2$]}&\\
    &model original&\multicolumn{3}{c}{with modification}&\\
    interaction &&\multicolumn{2}{c}{EPOS-based $R_1^{\mathrm{Ref}}$}&\multicolumn{2}{c}{SIBYLL-based $R_1^{\mathrm{Ref}}$}\\
    model &&&difference&&difference\\
    \hline
    EPOS-LHC&807.5$\pm$0.3&809.5$\pm$0.3&2.0$\pm$0.4 &803.3$\pm$0.3&-4.2$\pm$0.4\\
    QGSJET~II-04&792.2$\pm$0.3&796.6$\pm$0.3&4.4$\pm$0.4 &788.5$\pm$0.3&-3.7$\pm$0.4\\
    SIBYLL~2.3c&819.6$\pm$0.3&828.5$\pm$0.3&8.9$\pm$0.4 &820.5$\pm$0.3&0.9$\pm$0.4 \\
    \hline
    maximum difference between models&27.4$\pm$0.4 &31.9$\pm$0.5& &32.0$\pm$0.4&\\
    \end{tabular}
    \normalsize
\end{table}

\begin{table}[!h]
    \centering
    \caption{Results of $\langle X_{\mathrm{max}}^{\mu}\rangle $ with cross-sectional fraction modifications over the whole air shower. $R_1^{\mathrm{Ref}}$ for the modification is set by fitting the ratio of the predictions by EPOS-LHC or SIBYLL~2.3c with energy above $10^{15}$~eV.}
    \label{tab:resamplingMu}
    \footnotesize
    \begin{tabular}{l||c|cc|cc}
    \hline
    &\multicolumn{4}{c}{$\langle X_{\mathrm{max}}^{\mu}\rangle $ [g/cm$^2$]}&\\
    &model original&\multicolumn{3}{c}{with modification}&\\
    interaction &&\multicolumn{2}{c}{EPOS-based $R_1^{\mathrm{Ref}}$ }&\multicolumn{2}{c}{SIBYLL-based case $R_1^{\mathrm{Ref}}$}\\
    model &&&difference&&difference\\
    \hline
    EPOS-LHC&590.5$\pm$0.3&591.4$\pm$0.3&0.9$\pm$0.4&586.1$\pm$0.3&-4.4$\pm$0.4\\
    QGSJET~II-04&552.5$\pm$0.3&558.6$\pm$0.3&6.1$\pm$0.4&548.5$\pm$0.3&-4.0$\pm$0.4\\
    SIBYLL~2.3c&579.6$\pm$0.3&589.0$\pm$0.3&9.4$\pm$0.4&580.6$\pm$0.3&1.0$\pm$0.4 \\
    \hline
    maximum difference between models&38.0$\pm$0.4&32.8$\pm$0.4& &37.6$\pm$0.4&\\
    \end{tabular}
    \normalsize
\end{table}

\subsection{\label{sec:XmaxDiffMass}Effects of diffractive collision modeling on $\langle X_{\mathrm{max}}\rangle $ and $\langle X_{\mathrm{max}}^{\mu}\rangle $}

\begin{figure}
    \centering
    \includegraphics[clip, width=0.9\columnwidth]{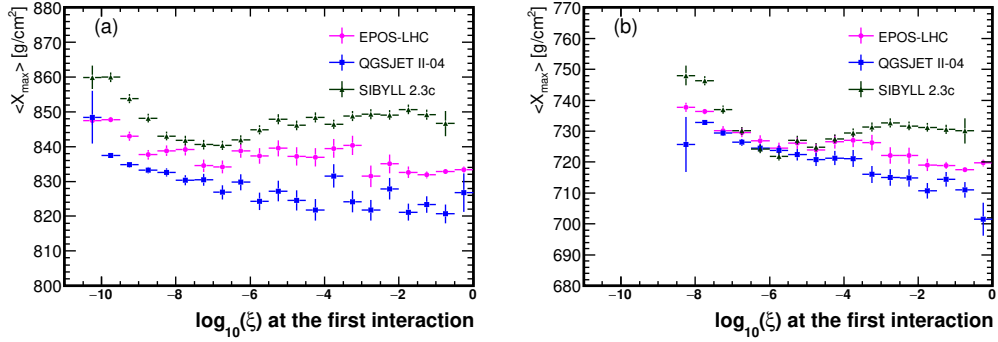}
    \caption{Diffractive-mass dependencies of $\langle X_{\mathrm{max}}\rangle $ for events whose collision type at the first interaction is the pSD collisions for (a) $10^{19}$~eV  and for (b) $10^{17}$~eV. }
    \label{fig:XmaxDiffMass}
\end{figure}

\begin{figure}
    \centering
    \includegraphics[clip, width=0.9\columnwidth]{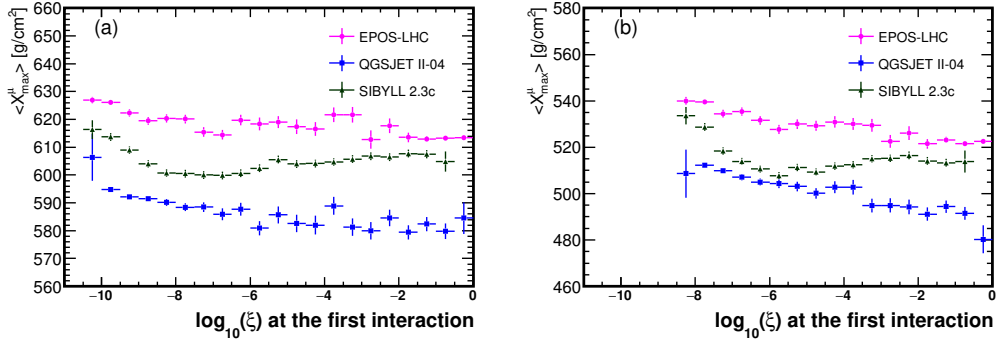}
    \caption{
    Diffractive-mass dependencies of $\langle X_{\mathrm{max}}^{\mu}\rangle $ for events whose collision type at the first interaction is the pSD collisions for (a) $10^{19}$~eV  and (b) $10^{17}$~eV. 
}
    \label{fig:XmaxMuDiffMass}
\end{figure}

As we discussed in Sec.~\ref{sec:XmaxFeature}, 
there are differences between the predictions of each model with the same collision type at the first interaction. These differences between models vary with the collision type. 
This suggests that the modeling of the diffractive collision affect $\langle X_{\mathrm{max}}\rangle $ and $\langle X_{\mathrm{max}}^{\mu}\rangle $. The remaining characteristics, the diffractive-mass spectrum and the particle productions of diffractive dissociation, may affect $\langle X_{\mathrm{max}}\rangle $ and $\langle X_{\mathrm{max}}^{\mu}\rangle $.
If the fraction of the low diffractive-mass events increases for the pSD collisions, the energies of particles become higher, thus the predictions of $\langle X_{\mathrm{max}}\rangle $ and $\langle X_{\mathrm{max}}^{\mu}\rangle $ for the pSD collision category, $\langle X_{\mathrm{max}}^{\mathrm{pSD}}\rangle $ and $\langle X_{\mathrm{max}}^{\mu, \mathrm{pSD}}\rangle $, become larger. 
Firstly, to understand the overview of the effect of the modeling, the diffractive-mass dependencies of $\langle X_{\mathrm{max}}^{ \mathrm{pSD}}\rangle $ and $\langle X_{\mathrm{max}}^{\mu,\, \mathrm{pSD}}\rangle $ are discussed. %
Figures~\ref{fig:XmaxDiffMass}~(a) and \ref{fig:XmaxMuDiffMass}~(a) display the diffractive-mass dependent $\langle X_{\mathrm{max}}^{ \mathrm{pSD}}\rangle $ and $\langle X_{\mathrm{max}}^{\mu,\, \mathrm{pSD}}\rangle $, respectively, for $10^{19}$~eV incident protons. The differences betwen models in Figures~\ref{fig:XmaxDiffMass}~(a) and \ref{fig:XmaxMuDiffMass}~(a) represent the differences with the same diffractive-mass pSD collisions at the first interactions for each models, thus the differences are originated from particle productions from diffractive dissociation and in interactions of secondary particles. 
The dependencies of $\langle X_{\mathrm{max}}^{\mathrm{pSD}}\rangle $ and $\langle X_{\mathrm{max}}^{\mu, \, \mathrm{pSD}}\rangle $ are similar; they become larger as the diffractive mass becomes smaller, and the diffractive-mass dependencies are small for $\log_{10}(\xi)>-6$.  

The difference between the maximum and the minimum $\langle X_{\mathrm{max}}^{\mathrm{pSD}}\rangle $ ($\langle X_{\mathrm{max}}^{\mu, \, \mathrm{pSD}}\rangle $) in the diffractive mass range is less than 30~$\mathrm{g/cm^2}$ in each model. This suggests that even if we assume an extreme diffractive-mass spectrum, a delta function for example, $\langle X_{\mathrm{max}}^{\mathrm{pSD}}\rangle $ and $\langle X_{\mathrm{max}}^{\mu,\, \mathrm{pSD}}\rangle $ are only affected by 30~$\mathrm{g/cm^2}$. Since projectile protons dissociate only for the pSD and the DD collisions, the diffractive-mass spectrum only affects the pSD and the DD collisions. 
In this section, firstly we discuss the effect on predictions of $\langle X_{\mathrm{max}}^{\mathrm{pSD}}\rangle $ and $\langle X_{\mathrm{max}}^{\mu, \mathrm{pSD}}\rangle $, then the effects on $\langle X_{\mathrm{max}}\rangle $ and $\langle X_{\mathrm{max}}^{\mu}\rangle $ are summarized later. 

It is difficult to separate the effects of particle productions from diffractive dissociation and the interactions of secondary particles. The difference between models for each diffractive-mass bin displayed in Figures~\ref{fig:XmaxDiffMass}~(a) and \ref{fig:XmaxMuDiffMass}~(a) reflects both the effects with the same diffractive mass.  
However, the differences in the diffractive-mass dependencies indicate the differences in particle productions from diffractive dissociation, because it is hard to explain the change of the diffractive-mass dependencies only using the effect of the interactions of secondary particles; if we assume the same particle productions from diffractive dissociation between SIBYLL~2.3c and EPOS-LHC, the interactions of secondary particles should make a few $\mathrm{g/cm^2}$ differences at $\log_{10}(\xi)=-8$ in Fig.~\ref{fig:XmaxDiffMass}~(a), whereas more than 15~$\mathrm{g/cm^2}$ differences is recognized at $\log_{10}(\xi)=-2$. 
The diffractive-mass dependencies of EPOS-LHC and QGSJET~II-04 are similar, whereas a difference can be seen between SIBYLL~2.3c and the other models; SIBYLL~2.3c only displays a small dip at $log_{10}(\xi)=-8$. These features suggest that particle productions from diffractive dissociation in SIBYLL~2.3c is different between the low and the middle diffractive-mass regions, and this feature affect $\langle X_{\mathrm{max}}^{\mathrm{pSD}}\rangle $ and $\langle X_{\mathrm{max}}^{\mu,\mathrm{pSD}}\rangle $.  
These diffractive-mass dependencies are same for $10^{17}$~eV proton primaries as displayed in Figures~\ref{fig:XmaxDiffMass}~(b) and \ref{fig:XmaxMuDiffMass}~(b). 
It is interesting that the models exhibit a good agreement in $\langle X_{\mathrm{max}}^{\mathrm{pSD}}\rangle $ predictions only around $\log_{10}(\xi)=-6$ as found in Fig.~\ref{fig:XmaxDiffMass}~(b).
The reason of such differences in diffractive-mass dependencies and energy dependence of the differences between models is worth studying in future.

To estimate the effect of the diffractive-mass spectrum quantitatively, the diffractive-mass spectrum at the first interaction was artificially changed.
$\langle X_{\mathrm{max}}^{\mathrm{pSD}}\rangle $  ($\langle X_{\mathrm{max}}^{\mu, \mathrm{pSD}}\rangle $) after modification of the diffrative mass spectrum is defined as, 
\begin{equation}
\langle X_{\mathrm{max}}^{({\mu})\mathrm{pSD},\,  \mathrm{modified}}\rangle = \sum_{i} P_{M_X}^{i} \langle X_{\mathrm{max}}^{({\mu})\mathrm{pSD}, \, i}\rangle ,
\label{eq:DiffMassModXmax}
\end{equation}
where $P_{M_X}^{i}$ is a probability of the $i$-th diffractive-mass bin of an arbitrary model in Fig.~\ref{fig:DiffMassSpectrum}, and  $\langle X_{\mathrm{max}}^{({\mu})\mathrm{pSD}, \, i}\rangle $ is $\langle X_{\mathrm{max}}^{(\mu) \mathrm{pSD}}\rangle $ taken from the $i$-th diffractive-mass bin of Fig.~\ref{fig:XmaxDiffMass}  (Fig.~\ref{fig:XmaxMuDiffMass}). Using this method, we can estimate the effect when the diffractive-mass spectrum are changed at the first interactions; if the cross-section of low diffractive-mass events are increased in the diffractive-mass spectrum, $\langle X_{\mathrm{max}}^{\mathrm{pSD},\, \mathrm{modified}}\rangle $ becomes larger. 
The effects of replacing $P_{M_X}^{i}$ between the three models on $\langle X_{\mathrm{max}}\rangle $ and $\langle X_{\mathrm{max}}^{\mu}\rangle $, $\mathrm{E}_{X_{\mathrm{max}}}$, are defined as follows,
\begin{equation}
    \mathrm{E}_{X_{\mathrm{max}}} = \langle X_{\mathrm{max}}^{({\mu})\mathrm{pSD},\, \mathrm{modified}}\rangle _{\mathrm{max}}- \langle X_{\mathrm{max}}^{({\mu})\mathrm{pSD}, \, \mathrm{modified}}\rangle _{\mathrm{min}},
    \label{eq:XmaxDelta}
\end{equation}
where $\langle X_{\mathrm{max}}^{({\mu})\mathrm{pSD}, \, \mathrm{modified}}\rangle _{\mathrm{max}}$ ($\langle X_{\mathrm{max}}^{({\mu})\mathrm{pSD}, \, \mathrm{modified}}\rangle _{\mathrm{min}}$) is the maximum (minimum) value of $\langle X_{\mathrm{max}}^{({\mu})\mathrm{pSD}, \, \mathrm{modified}}\rangle $ in the three models.
%

The results of $\langle X_{\mathrm{max}}^{\mathrm{pSD}, \, \mathrm{modified}}\rangle $ are displayed in Tab.~\ref{tab:Xmax_changeDiffmass} for $10^{19}$~eV and in Tab.~\ref{tab:Xmax_changeDiffmass_17} for $10^{17}$~eV. If the diffractive-mass spectrum of QGSJET~II-04 is replaced by a spectrum of another model, $\langle X_{\mathrm{max}}^{\mathrm{pSD}, \, \mathrm{modified}}\rangle $ becomes smaller. %
The size of $\mathrm{E}_{X_{\mathrm{max}}}$ is (4.1$\pm$0.7) to (5.1$\pm$0.7)~g/cm$^2$ for $10^{19}$~eV and (4.3$\pm$0.7) to (7.1$\pm$1.2)~g/cm$^2$ for $10^{17}$~eV. 
Differences between the model predictions when the same diffractive-mass spectrum was assumed at the first interaction are summarized in the bottom row of Tab.~\ref{tab:Xmax_changeDiffmass} and Tab.~\ref{tab:Xmax_changeDiffmass_17}. 
The size of the differences is (18.6$\pm$0.7) to (21.0$\pm$1.2)~$\mathrm{g/cm^2}$ for $10^{19}$~eV, which is larger by 4.5-6.9~$\mathrm{g/cm^2}$ than the original difference.
Therefore, the differences of predictions between models become larger even when the same diffractive-mass spectrum is used in the first pSD collisions. 
It is noted that these differences include the effect of particle productions from diffractive dissociation and the interactions of secondary particles, and these two effects cannot be separated. 

The results of $\langle X_{\mathrm{max}}^{\mu,\mathrm{pSD}, \, \mathrm{modified}}\rangle $ are displayed in Tab.~\ref{tab:XmaxMu_changeDiffmass} for $10^{19}$~eV and in Tab.~\ref{tab:XmaxMu_changeDiffmass_17} for $10^{17}$~eV. The tendencies of the effect of the diffractive-mass spectrum are similar as the $\langle X_{\mathrm{max}}^{\mathrm{pSD}, \, \mathrm{modified}}\rangle $ case; if the diffractive-mass spectrum of QGSJET~II-04 is replaced by that in another model, $\langle X_{\mathrm{max}}^{\mu, \mathrm{pSD}, \, \mathrm{modified}}\rangle $ becomes smaller. %
$\mathrm{E}_{X_{\mathrm{max}}}$ is (2.8$\pm$0.7) to (4.8$\pm$0.7)~g/cm$^2$ for $10^{19}$ eV and (4.4$\pm$0.8) to (7.1$\pm$0.8)~g/cm$^2$ for $10^{17}$~eV.   
The differences between the model predictions become larger even when the same diffractive-mass spectrum is used in the first pSD collisions.

The estimation above is only focused on the pSD collision category. The diffractive-mass spectrum is only related to particle productions in the dissociation system of the diffractive collision, and they mainly affect the pSD and the DD collision categories. 
Because the cross-section of the pSD and the DD collisions are 5 to 12~\% of the inelastic collisions, the effect of the diffractive collision modeling at the first interactions on $\langle X_{\mathrm{max}}\rangle $ and $\langle X_{\mathrm{max}}^{\mu}\rangle $ is approximately 10 times smaller than the estimations above.
Therefore, the effects of the diffractive-mass spectrum on the average of predictions of $\langle X_{\mathrm{max}}\rangle $ and $\langle X_{\mathrm{max}}^{\mu}\rangle $ are expected to be 0.5~$\mathrm{g/cm^2}$ at maximum. These effects of diffractive-mass spectrum are much smaller than the effect of the cross-sectional fractions, which are 5.0~$\mathrm{g/cm^2}$ at maximum as discussed in Sec.~\ref{sec:XmaxFractionFirstint}.  
The effects of the diffractive-mass spectrum are similar in $\langle X_{\mathrm{max}}\rangle $ and $\langle X_{\mathrm{max}}^{\mu}\rangle $. 

From the results in Sec.~\ref{sec:XmaxFraction} and \ref{sec:XmaxDiffMass}, the effects of the diffractive collisions on $\langle X_{\mathrm{max}}\rangle $ show similar size and the same direction with that on $\langle X^{\mu}_{\mathrm{max}}\rangle $. Therefore, 
both of the effects of the cross-sectional fraction and the diffractive mass spectrum do not solve the problem of inconsistent interpretations on the cosmic-ray mass composition estimated from $\langle X_{\mathrm{max}}\rangle $ and $\langle X_{\mathrm{max}}^{\mu}\rangle $. It must be noted that the collisions of low energy projectiles below $10^{15}$~eV also affect $\langle X_{\mathrm{max}}\rangle $ and $\langle X_{\mathrm{max}}^{\mu}\rangle $. According to Ref.~\cite{Ostapchenko_2016}, collisions with lower energy projectiles affect more on $\langle X_{\mathrm{max}}^{\mu}\rangle $ than $\langle X_{\mathrm{max}}\rangle $.

\begin{table}[!h]
    \centering
    \caption{The results of $\langle X_{\mathrm{max}}^{\mathrm{pSD}}\rangle $ with the modification by the diffractive-mass spectrum, $\langle X_{\mathrm{max}}^{\mathrm{pSD}, \, \mathrm{modified}}\rangle $, for the $10^{19}$~eV proton primary case. $\mathrm{E}_{X_{\mathrm{max}}}$ is the size of effects estimated by three diffractive-mass spectra as defined in Equation~\ref{eq:XmaxDelta}. }
    \label{tab:Xmax_changeDiffmass}
    \footnotesize
    \begin{tabular}{l||c|ccc|c}
    \hline
    &\multicolumn{3}{c}{$\langle X_{\mathrm{max}}^{\mathrm{pSD}, \, \mathrm{modified}}\rangle $ [g/cm$^2$]}&&\\
    &&\multicolumn{2}{c}{model for a diffractive-mass spectrum} & & the size of the effects\\
    model &original model &EPOS-LHC&QGSJET~II-04&SIBYLL~2.3c&$\mathrm{E}_{X_{\mathrm{max}}}$ [g/cm$^2$]\\
    \hline
    EPOS-LHC& 838.0$\pm$0.3& 838.0$\pm$0.3 & 841.7$\pm$0.5 & 837.6$\pm$0.5 & 4.1$\pm$0.7\\
    QGSJET~II-04&833.3 $\pm$0.3& 829.3$\pm$1.0 & 833.3$\pm$0.3 & 828.2$\pm$0.6 & 5.1$\pm$0.7\\
    SIBYLL~2.3c&847.4$\pm$0.3& 850.3$\pm$0.7&851.9$\pm$0.6& 847.4$\pm$0.3& 4.4$\pm$0.7\\
    \hline
    maximum difference&14.1$\pm$0.5&21.0$\pm$1.2&18.6$\pm$0.7&19.2$\pm$0.6&\\
    \end{tabular}
    \normalsize
\end{table}

\begin{table}[!h]
    \centering
    \caption{The results of $\langle X_{\mathrm{max}}^{\mathrm{pSD}, \, \mathrm{modified}}\rangle $ for the $10^{17}$~eV proton primary case.  }
    \label{tab:Xmax_changeDiffmass_17}
    \footnotesize
    \begin{tabular}{l||c|ccc|c}
    \hline
    &\multicolumn{3}{c}{$\langle X_{\mathrm{max}}^{\mathrm{pSD}, \, \mathrm{modified}}\rangle $ [g/cm$^2$]}&&\\
    &&\multicolumn{2}{c}{model for a diffractive-mass spectrum} & & the size of the effects\\
    model &original model&EPOS-LHC&QGSJET~II-04&SIBYLL~2.3c&$\mathrm{E}_{X_{\mathrm{max}}}$ [g/cm$^2$]\\
    \hline
    EPOS-LHC&725.5 $\pm$0.3& 725.5$\pm$0.4& 729.8$\pm$0.5 & 725.9$\pm$0.5 & 4.3$\pm$0.7\\
    QGSJET~II-04&726.5$\pm$0.4& 719.4$\pm$1.0 & 726.5$\pm$0.4 & 720.9$\pm$0.6 & 7.1$\pm$1.2\\
    SIBYLL~2.3c&730.6$\pm$0.4& 734.7$\pm$0.7&735.5$\pm$0.6& 730.7$\pm$0.4& 4.8$\pm$0.7\\
    \hline
    maximum difference&5.2$\pm$0.5&15.3$\pm$1.2&9.0$\pm$0.7&9.8$\pm$0.7&\\
    \end{tabular}
    \normalsize
\end{table}

\begin{table}[!h]
    \centering
    \caption{The results of $\langle X_{\mathrm{max}}^{\mu, \mathrm{pSD}, \, \mathrm{modified}}\rangle $ for the $10^{19}$ eV proton primary case. 
    }
    \label{tab:XmaxMu_changeDiffmass}
    \footnotesize
    \begin{tabular}{l||c|ccc|c}
    \hline
    &\multicolumn{3}{c}{$\langle X_{\mathrm{max}}^{\mu, \mathrm{pSD}, \, \mathrm{modified}}\rangle $ [g/cm$^2$]}&&\\
    &&\multicolumn{2}{c}{model for a diffractive-mass spectrum} & & the size of the effects\\
    model &original model&EPOS-LHC&QGSJET~II-04&SIBYLL~2.3c&$\mathrm{E}_{X_{\mathrm{max}}}$ [g/cm$^2$]\\
    \hline
    EPOS-LHC& 618.2 $\pm$0.3& 618.2$\pm$0.3& 621.7$\pm$0.5& 618.4$\pm$0.5& 3.5$\pm$0.6\\
    QGSJET~II-04&590.9$\pm$0.3& 587.4$\pm$1.1& 591.0$\pm$0.3& 586.2$\pm$0.6& 4.8$\pm$0.7\\
    SIBYLL~2.3c&604.8$\pm$0.3& 607.1$\pm$0.7&607.6$\pm$0.6& 604.8$\pm$0.3& 2.8$\pm$0.7\\
    \hline
    maximum difference&27.3$\pm$0.5&30.9$\pm$1.3&30.7$\pm$0.7&32.2$\pm$0.7&\\
    \end{tabular}
    \normalsize
\end{table}

\begin{table}[!h]
    \centering
    \caption{The results of $\langle X_{\mathrm{max}}^{\mu, \mathrm{pSD}, \, \mathrm{modified}}\rangle $ for the $10^{17}$ eV proton primary case.  }
    \label{tab:XmaxMu_changeDiffmass_17}
    \footnotesize
    \begin{tabular}{l||c|ccc|c}
    \hline
    &\multicolumn{3}{c}{$\langle X_{\mathrm{max}}^{\mu, \mathrm{pSD}, \, \mathrm{modified}}\rangle $ [g/cm$^2$]}&&\\
    &&\multicolumn{2}{c}{model for a diffractive-mass spectrum} & & the size of the effects\\
    model &model original&EPOS-LHC&QGSJET~II-04&SIBYLL~2.3c&$\mathrm{E}_{X_{\mathrm{max}}}$ [g/cm$^2$]\\
    \hline
    EPOS-LHC&529.1$\pm$0.4& 529.1$\pm$0.4& 533.8$\pm$0.6& 529.5$\pm$0.6& 4.7$\pm$0.7\\
    QGSJET~II-04&506.6$\pm$0.4& 499.4$\pm$1.1& 506.6$\pm$0.4& 501.1$\pm$0.6& 7.1$\pm$0.8\\
    SIBYLL~2.3c&514.5$\pm$0.4& 518.6$\pm$0.8&518.8$\pm$0.7& 514.4$\pm$0.4& 4.4$\pm$0.8\\
    \hline
    maximum difference&22.5$\pm$0.6&29.7$\pm$1.2&27.3$\pm$0.7&28.4$\pm$0.9&\\
    \end{tabular}
    \normalsize
\end{table}

\section{\label{sec:SigmaXmax}The fluctuation of $X_{\mathrm{max}}$}
The fluctuation of $X_{\mathrm{max}}$, $\sigma(X_{\mathrm{max}})$, is another mass sensitive observable widely used in experiments. 
In this section, we estimate the effect of the ratio of diffractive collisions $R_1$, which shows the largest effects on $\langle X_{\mathrm{max}}\rangle $ as discussed in Sec. 4, on $\sigma(X_{\mathrm{max}})$ for one simple case. 

Table~\ref{tab:SigmaXmax} shows $\sigma(X_{\mathrm{max}})$ computed as the standard deviation of each sample categorized at the first interaction for the $10^{19}$~eV proton primary case. The $\sigma(X_{\mathrm{max}})$ values for all categories of diffractive collisions are slightly larger than that of the non-diffracitve category. One should note that large differences in $\langle X_{\mathrm{max}}\rangle $ and fractions should be taken into account when we consider $\sigma(X_{\mathrm{max}})$; 
the categories of diffractive collisions show approximately 40-50~$\mathrm{g/cm^2}$ larger $\langle X_{\mathrm{max}}\rangle $ than that of the average, and the fraction of diffractive collisions is 0.074 to 0.182 as shown in Figures~\ref{fig:FractionAtFirstInt} and \ref{fig:XmaxWithCategory}. Because of the large difference of the mean values between ND and the others, the diffractive components affect on the $\sigma(X_{\mathrm{max}})$ even the fraction is small. 
\begin{table}[!h]
    \centering
    \caption{$\sigma(X_{\mathrm{max}}) $ with categorization at the first interaction.}
    \label{tab:SigmaXmax}
    \footnotesize
    \begin{tabular}{l|ccccc|c}
    \hline
    & ND&pSD&tSD&DD&CD&original\\
    \hline
    EPOS-LHC&53.8&65.3&69.5&61.8&66.0&58.2\\	
    QGSJET II-04&57.1&66.5&76.3&66.8&92.6&60.5\\
    SIBYLL~2.3c&60.5&64.2&76.9&62.6& &61.8\\
    \end{tabular}
    \normalsize
\end{table}

The estimation of the effect on $\sigma(X_{\mathrm{max}})$ at the first interaction is performed by calculating the $X_{\mathrm{max}}$ distribution with changing $R_1$.
The $X_{\mathrm{max}}$ distribution of each category is re-scaled according to the modified fraction obtained from Eq.~\ref{eq:fraction}, and the modified $\sigma(X_{\mathrm{max}})$ is calculated from the combined $X_{\mathrm{max}}$ distribution of them.
The results of $\sigma(X_{\mathrm{max}})$ at the minimum and maximum $R_1$ values, which is $R_1 = 0.074$ and $R_1=0.182$, respectively, are shown in Table~\ref{tab:ModifySigmaXmax}. 
The size of the effects when $R_1$ is changed from 0.074 to 0.182 is 1.6~$\mathrm{g/cm^2}$ at SIBYLL~2.3c and 2.8~$\mathrm{g/cm^2}$ at QGSJET~II-04. 
Moreover, if the same $R_1$ is used, the differences between models become larger than the original differences between models; the size of differences between models with the same $R_1$ is 5.2~$\mathrm{g/cm^2}$ for the maximum $R_1$ case and 6.1~$\mathrm{g/cm^2}$ for the minimum $R_1$ case, while the original difference is 2.6~$\mathrm{g/cm^2}$. 
It seems that the effects of diffractive collisions are canceled out with the differences in other parts by chance. 
\begin{table}[!h]
    \centering
    \caption{Results of $\sigma(X_{\mathrm{max}}) $ with modifications of $R_1$ at the first interaction.}
    \label{tab:ModifySigmaXmax}
    \footnotesize
    \begin{tabular}{l||c|cc|cc}
    \hline
    &\multicolumn{4}{c}{$\sigma(X_{\mathrm{max}})$ [g/cm$^2$]}&\\
    &model original&\multicolumn{3}{c}{with modification}&\\
    interaction &&\multicolumn{2}{c}{ $R_1 =0.182$}&\multicolumn{2}{c}{$R_1 = 0.074$}\\
    model &&&difference&&difference\\
    \hline
    EPOS-LHC&58.2&58.2&0.0 &55.7&-2.5\\
    QGSJET~II-04&60.5&62.2&1.7 &59.4&-1.1\\
    SIBYLL~2.3c&61.8&63.4&1.6 &61.8&0.0\\
    \hline
    maximum difference between models&2.6 &5.2& &6.1&\\
    \end{tabular}
    \normalsize
\end{table}

The estimation of effects over the whole air shower is performed following the method used in section~\ref{sec:XmaxFractionResampling}. The results of $\sigma(X_{\mathrm{max}})$ are shown in Table~\ref{tab:resamplingSigmaXmax}. The size of the effects is 2.4-3.0~$\mathrm{g/cm^2}$. Comparing the estimation at the first interaction, the size of the effects is similar for EPOS-LHC and 0.9~$\mathrm{g/cm^2}$ larger for QGSJET II-04 and SIBYLL 2.3c. 
We confirmed that the first interaction have a dominant contoribution on $\sigma(X_{\mathrm{max}})$.
\begin{table}[!h]
    \centering
    \caption{Results of $\sigma(X_{\mathrm{max}}) $ with cross-section ratio modifications over the whole air shower. The reference ratio, $R_1^{\mathrm{Ref}}$, for the modification is set by fitting the ratio of the predictions by EPOS-LHC or SIBYLL~2.3c with energy above $10^{15}$~eV.}
    \label{tab:resamplingSigmaXmax}
    \footnotesize
    \begin{tabular}{l||c|cc|cc}
    \hline
    &\multicolumn{4}{c}{$\sigma(X_{\mathrm{max}})$ [g/cm$^2$]}&\\
    &model original&\multicolumn{3}{c}{with modification}&\\
    interaction &&\multicolumn{2}{c}{EPOS-based $R_1^{\mathrm{Ref}}$}&\multicolumn{2}{c}{SIBYLL-based $R_1^{\mathrm{Ref}}$}\\
    model &&&difference&&difference\\
    \hline
    EPOS-LHC&58.2&58.0&0.2 &55.6&-2.6\\
    QGSJET~II-04&60.5&63.1&2.6 &59.4&-1.1\\
    SIBYLL~2.3c&61.8&64.8&3.0 &62.3&0.5\\
    \hline
    maximum difference between models&2.6 &6.8& &6.7&\\
    \end{tabular}
    \normalsize
\end{table}

The size of effects of $R_1$ on $\sigma(X_{\mathrm{max}})$ is 3.0~$\mathrm{g/cm^2}$ at maximum, which is larger than the original differences of $\sigma(X_{\mathrm{max}})$ between models. However, the effect is smaller than the difference of $\sigma(X_{\mathrm{max}})$ between proton (approximately 60~$\mathrm{g/cm^2}$) and helium primaries (approximately 40~$\mathrm{g/cm^2}$); 
therefore the effect of $R_1$ makes no significant impact on the interpretation of the mass composition from $\sigma(X_{\mathrm{max}})$. 

\section{\label{sec:Conclusion}Discussion and Conclusion}

In this work, the effects of the diffractive collisions on $\langle X_{\mathrm{max}}\rangle $, $\sigma(X_{\mathrm{max}})$, and $\langle X_{\mathrm{max}}^{\mu}\rangle $ were studied with focus on three detail characteristics of the diffractive collisions, (1) cross-sectional fractions among collision types, (2) diffractive-mass spectra, and (3) particle productions from diffractive dissociation. 
The diffractive collisions make $\langle X_{\mathrm{max}}\rangle $ and $\langle X_{\mathrm{max}}^{\mu}\rangle $ larger, and the cross-sectional fraction of the diffractive collisions in the inelastic collisions displays the largest effect among the detail characteristics. 

If we assume the current differences in predictions between models as uncertainties, the maximum effect of $R_1$ over a whole air shower is 8.9$\pm$0.4~$\mathrm{g/cm^2}$ and 9.4$\pm$0.4~$\mathrm{g/cm^2}$ for $\langle X_{\mathrm{max}}\rangle $ and $\langle X_{\mathrm{max}}^{\mu}\rangle $, respectively. If the same cross-sectional fraction of the diffractive collisions is used for collisions of $>10^{15}$~eV projectiles, the differences between the model predictions become larger by approximately 4.5~$\mathrm{g/cm^2}$ for $\langle X_{\mathrm{max}}\rangle $ and smaller by 0.4-5.2~$\mathrm{g/cm^2}$ for $\langle X_{\mathrm{max}}^{\mu}\rangle $, therefore the differences of interpretation of the mass composition from $\langle X_{\mathrm{max}}\rangle $ between models become larger.  
The effect of $R_1$ at the first interaction on $\langle X_{\mathrm{max}}\rangle $ and $\langle X_{\mathrm{max}}^{\mu}\rangle $ is approximately 5~$\mathrm{g/cm^2}$ for $10^{19}$~eV, whereas the effects of other cross-sectional fractions and the diffractive-mass spectrum at the first interaction are less than 1~$\mathrm{g/cm^2}$ for $10^{19}$~eV. Therefore, the effects of cross-sectional fractions between the pSD, tSD, and DD collisions and the diffractive-mass spectrum are negligible for both $\langle X_{\mathrm{max}}\rangle $ and $\langle X_{\mathrm{max}}^{\mu}\rangle $. 
Any details of diffractive collisions discussed in this paper cannot solve the differences of $\langle X_{\mathrm{max}}\rangle $ and $\langle X_{\mathrm{max}}^{\mu}\rangle $ predictions between the models. The effect of $R_1$ on $\sigma(X_{\mathrm{max}})$ is also discussed, however, the effect makes no significant impact on the mass composition from $\sigma(X_{\mathrm{max}})$.

We found that the sizes of the effect on $\langle X_{\mathrm{max}}\rangle $ and $\langle X_{\mathrm{max}}^{\mu}\rangle $ by any details of diffractive collisions are similar.
This suggests that the discussions in this paper cannot explain the discrepancy between the interpretations of the cosmic-ray mass composition from $\langle X_{\mathrm{max}}\rangle $ and $\langle X_{\mathrm{max}}^{\mu}\rangle $.
The effect of low energy collisions of $< 10^{15}$~eV projectiles are not discussed in this paper, however, they will affect more on $\langle X_{\mathrm{max}}^{\mu}\rangle $ than on $\langle X_{\mathrm{max}}\rangle $~\cite{Ostapchenko_2016}. 
Thus, it is important to study the effect of low energy collisions carefully in the future. %

The effects of particle productions from diffractive dissociation were discussed when a difference was found after assuming the same collision type and the same diffractive-mass spectrum at the first interaction. 
If the same diffractive-mass spectrum is used for the pSD collisions at the first interaction, the differences between the model predictions for $\langle X_{\mathrm{max}}^{\mathrm{pSD}}\rangle $ and $\langle X_{\mathrm{max}}^{\mu, \mathrm{pSD}}\rangle $ are 21.0$\pm$1.2~$\mathrm{g/cm^2}$ and 30.9$\pm$1.3~$\mathrm{g/cm^2}$ at maximum for $10^{19}$~eV, respectively. It is noted that the effect of these differences on $\langle X_{\mathrm{max}}\rangle $ and $\langle X_{\mathrm{max}}^{\mu}\rangle $ at the first interaction is ten times smaller than written above due to the small fraction of the pSD and the DD collisions. These differences are caused by two effects, the effects of particle productions from diffractive dissociation and the interactions of secondary particles, which cannot be separated in this study. 
The effect of particle productions from diffractive dissociation is worth studying in the future.

Finally, the importance of improvements of hadronic interaction models and the experiments are discussed. 
The improvements of the cross-sectional fraction of the diffractive collisions in the inelastic collisions are most important since it exhibits the large effect on the mean value of $\langle X_{\mathrm{max}}\rangle $ and $\langle X_{\mathrm{max}}^{\mu}\rangle $. 
If we focus on collision types in diffractive collisions, there are large differences in the cross-sectional fractions of the DD and the CD collisions between the models, while relatively small differences in that of the pSD and tSD collisions. Therefore the major sources of the differences of the cross-sectional fraction of diffractive collisions between the models are the DD and CD collisions. 
The latest models are updated using the parts of results in the LHC, but not integrating the latest results on the diffractive collisions at the LHC. It is important to update these models using the latest experimental results of the DD collisions. 
Another important point is low diffractive-mass regions of diffractive collisions. Experimentally, there is a difficulty in the measurements for the low diffractive-mass collisions, because most of the particles are produced in very forward regions around the beam pipes. There is a large peak of the cross-section at very low diffractive-mass regions in predictions by several models. Therefore, the experimental uncertainty of the cross-section of diffractive collisions is large due to that difficulty. Moreover, the cross sections of heavy ions collisions, {\it i.e.} lead and xenon collisions, are measured in the LHC, whereas that of the light ion collisions, which emulate the interactions between a cosmic-ray particle and an atmospheric nuclei are not measured so far. 
Thus, the measurements for low-diffractive mass regions and light ion collisions are required.
Recently, the ATLAS and the LHCf collaborations displayed the first results of their joint analysis for very forward photons produced by the diffractive collisions~\cite{ATLAS-CONF-2017-075}, thanks to their complementary pseudorapidity coverage sharing the same interaction point at the LHC. The measurements by the ATLAS and the LHCf collaborations are helpful in improving the treatments of the low diffractive-mass collisions in the hadronic interaction models. 

\section*{Acknowledgements}
We thank N. Sakurai for providing the programs for resampling in CONEX. We are grateful to F. Riehn for useful discussions and comments and to C. Baus, T. Pierog, and R. Ulrich for the implementation of the CRMC interface tool. This work was supported by the Japanese Society for the Promotion of Science (JSPS) KAKENHI (Grant Numbers 18H01227) in Japan and by the joint research program of the Institute for Cosmic Ray Research (ICRR), University of Tokyo.


\bibliographystyle{ptephy}
\bibliography{Author_tex}
%



\end{document}